\documentclass[journal]{IEEEtran}

\usepackage{graphicx}
\usepackage{subfig}
\usepackage{array}
\usepackage{lipsum}
\usepackage[font={footnotesize}]{caption}
\usepackage{bm}
\usepackage{amsmath}
\usepackage{upgreek}
\usepackage{physics}
\usepackage{esvect}
\usepackage{color, soul}
\usepackage{cite}
\newcolumntype{?}{!{\vrule width 1pt}}
\usepackage{amssymb}
\usepackage{dblfloatfix}
\usepackage{midfloat}

\begin{document}

\title{High-Scalability CMOS Quantum Magnetometer with Spin-State Excitation and Detection of Diamond Color Centers}
\author{Mohamed I. Ibrahim,~\IEEEmembership{Student Member,~IEEE},
		Christopher Foy,
			Dirk R. Englund~\IEEEmembership{Member,~IEEE},
        and~Ruonan~Han,~\IEEEmembership{Senior~Member,~IEEE}\vspace{-0.8cm}       
\thanks{This is the extended version of a paper originally presented at the IEEE Solid-State Circuit Conference (ISSCC), San Francisco, CA, USA, February~2019. This work is supported by National Science Foundation (NSF) Research Advanced by Interdisciplinary Science and Engineering (RAISE) Transformational Advances in Quantum Systems (TAQS) (Grant No. 1839159), MIT Center for Integrated Circuits \& Systems, Singapore-MIT Research Alliance (Low Energy Electronic Systems IRG), Army Research Office MURI on ``Imaging and Control of Biological Transduction using NV-Diamond", and Gordon $\&$ Betty Moore Foundation.}
\thanks{M. I. Ibrahim$^*$, C. Foy$^*$ ($^*$: equal contributors), D. R. Englund and R. Han are with the Department of Electrical Engineering and Computer Science, Massachusetts Institute of Technology, Cambridge, MA 02139, USA (email: ibrahimm@mit.edu; cfoy3@mit.edu; englund@mit.edu; ruonan@mit.edu).}
}
\maketitle

\begin{abstract}\label{sec_abstract}
Magnetometers based on quantum mechanical processes enable high sensitivity and long-term stability without the need for re-calibration, but their integration into fieldable devices remains challenging. This paper presents a CMOS quantum vector-field magnetometer that miniaturizes the conventional quantum sensing platforms using nitrogen-vacancy (NV) centers in diamond. By integrating key components for spin control and readout, the chip performs magnetometry through optically detected magnetic resonance (ODMR) through a diamond slab attached to a custom CMOS chip. The ODMR control is highly uniform across the NV centers in the diamond, which is enabled by a CMOS-generated $\sim$2.87~GHz magnetic field with $<$5$\%$ inhomogeneity across a large-area  current-driven wire array. The magnetometer chip is 1.5~mm$^2$ in size, prototyped in 65-nm bulk CMOS technology, and attached to a 300$\times$80~$\upmu$m$^2$ diamond slab. NV fluorescence is measured by CMOS-integrated photodetectors. This on-chip measurement is enabled by efficient rejection of the green pump light from the red fluorescence through a CMOS-integrated spectral filter based on a combination of spectrally dependent plasmonic losses and diffractive filtering  in the CMOS back-end-of-line (BEOL). This filter achieves $\sim$25~dB of green light rejection. We measure a sensitivity of 245~nT/Hz$^{1/2}$, marking a 130$\times$ improvement over a previous CMOS-NV sensor prototype, largely thanks to the better spectral filtering and homogeneous microwave generation over larger area.

\end{abstract}

\begin{IEEEkeywords}
CMOS, quantum, magnetometry, nitrogen-vacancy centers, Zeeman, nanophotonic filter, field homogeneity
\end{IEEEkeywords}

\section{Introduction}\label{sec_introduction}

Solid-state quantum sensors are attracting broad interests thanks to a combination of excellent sensitivity and long-term stability. In particular, ensembles of nitrogen-vacancy (NV) centers in diamond have emerged as an outstanding \emph{room-temperature} sensor platform \cite{Gruber1997ScanningCenters,Balasubramanian2008,Taylor2008High-SensitivityResolution,LeSage2013OpticalCells,Steinert2013MagneticResolution}. Ensemble NV-based magnetometry has achieved sensitivities at or below the picotesla level\cite{PhysRevX.5.041001,Clevenson2014}, with applications ranging from bacteria magnetic imaging\cite{LeSage2013OpticalCells}, NMR spectroscopy\cite{bucher2018hyperpolarizationenhanced} to wide-field microscopy of superconducting materials \cite{PhysRevApplied.10.034032}. However, an impediment to fieldable devices lies in the co-integration of different subsystems needed for optically detected magnetic resonance (ODMR), including microwave generation/delivery to diamond,  optical excitation, filtering and fluorescence-based spin detection (see Section~\ref{sec_physics}). 

We recently addressed this integration problem by performing ODMR with a custom-designed CMOS circuit coupled to a diamond NV ensemble\cite{Ibrahim2018}, \cite{Kim2019a}. However, insufficient pump light rejection and limited area of homogeneous microwave driving fields for the ODMR measurements posed a major limitation on achievable magnetic field sensitivity. Here, we present a new CMOS prototype that addresses these problems to achieve a $>$100$\times$ sensitivity improvement, down to 245 nT/Hz$^{1/2}$\cite{Ibrahim2019ACapability}. 

The manuscript is organized as follows. Section~\ref{sec_physics} summarizes the basics quantum magnetometry and reviews the first NV-CMOS prototype\cite{Ibrahim2018,Kim2019a}.  Section~\ref{sec_arch} describes the improved chip architecture that incorporates high-homogeneity microwave delivery and diffraction- and plasmonic optical filtering, realized in a 65-nm CMOS process. Section~\ref{sec_measurement} presents performance measurements. Section~\ref{sec_conclusion} concludes  with an outlook on further improvements to magnetometry and the incorporation of additional quantum sensing functions. 
\section{Background and Prior Research}\label{sec_physics}


\begin{figure}
\centering
\subfloat[]{\includegraphics[width=3.4in]{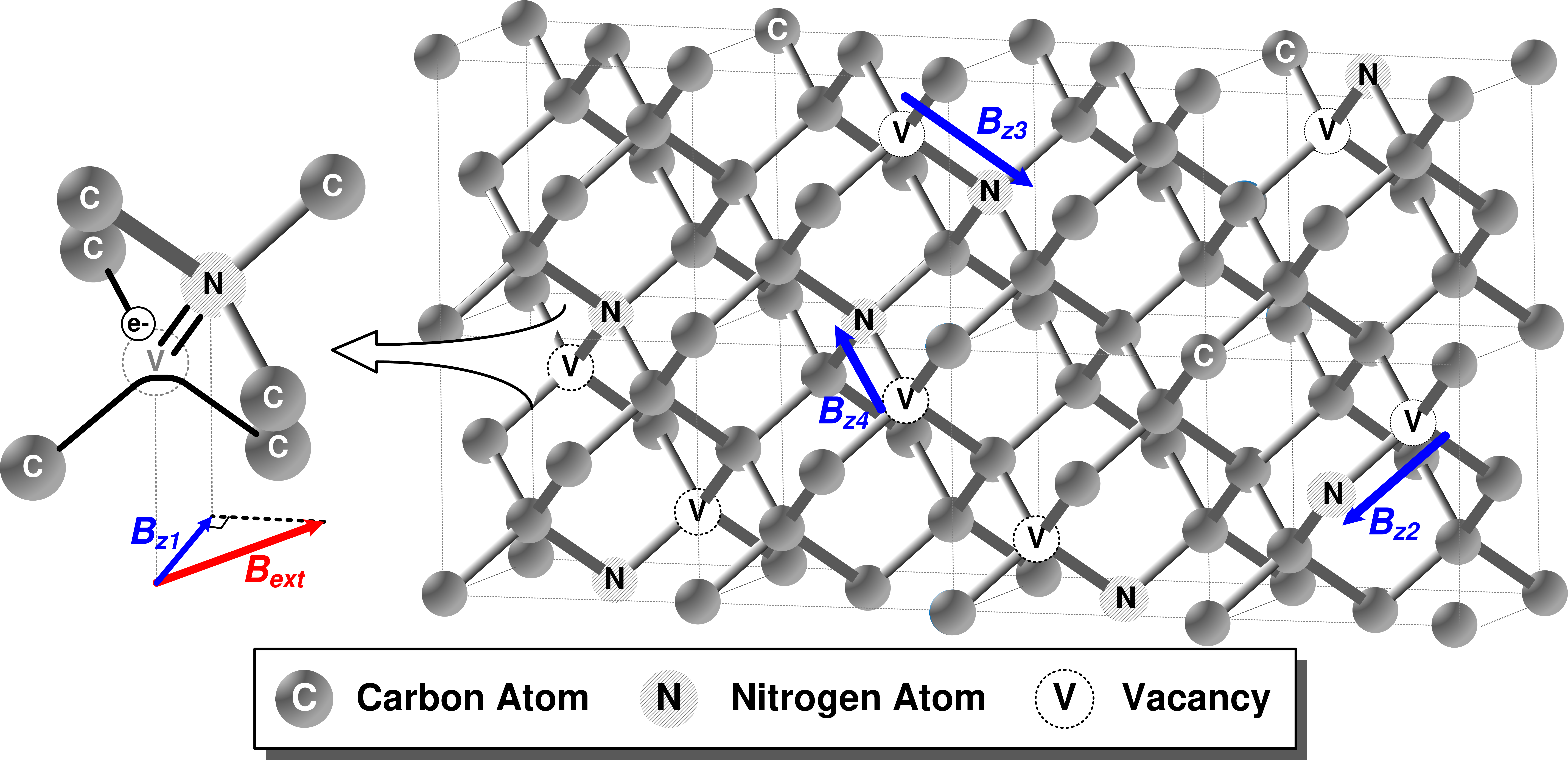}\label{fig_nv_centers}}\\
\subfloat[]{\includegraphics[width=2.8in]{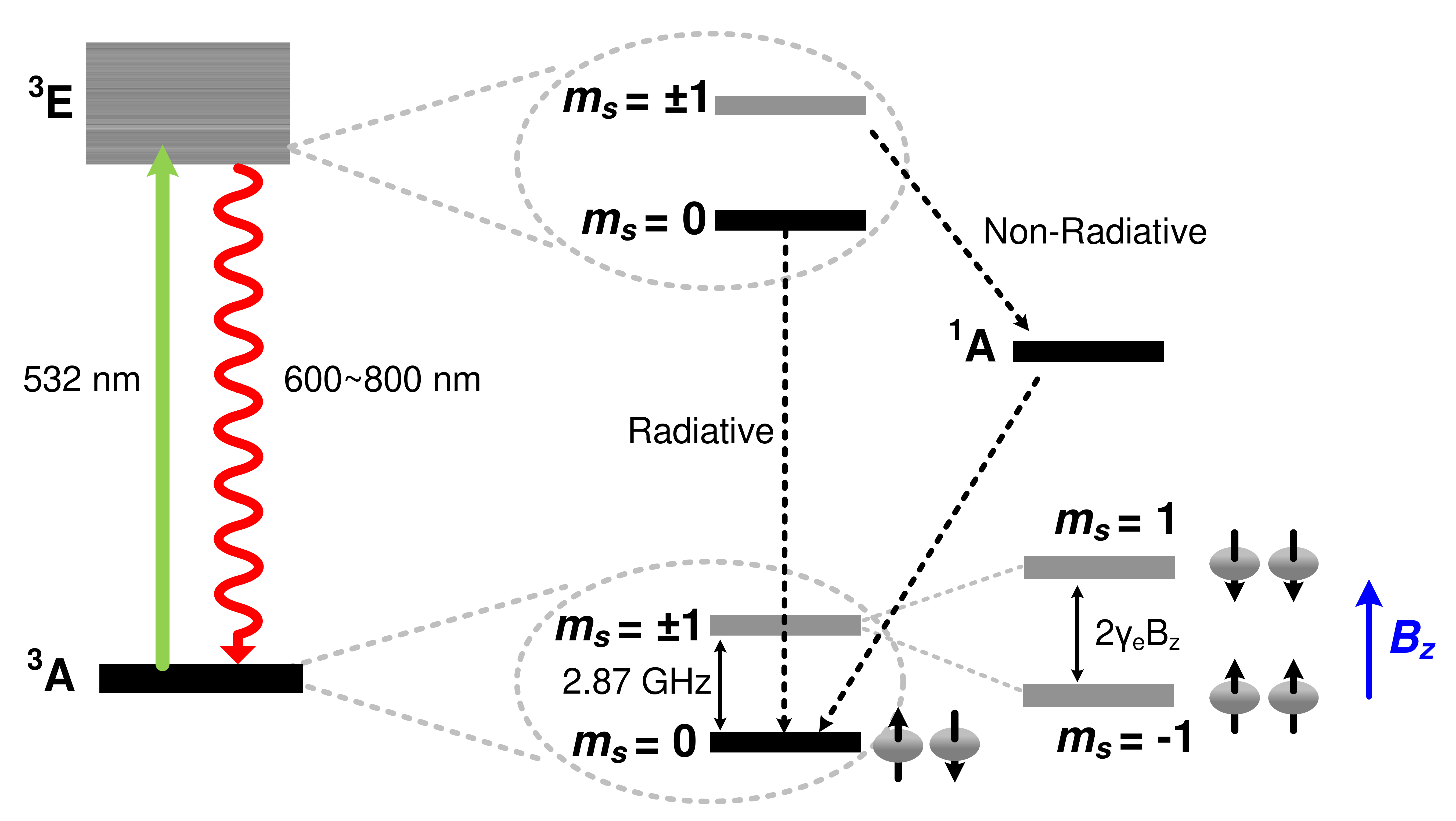}\label{fig_nv_energy}}
\caption{(a) Negatively-charged nitrogen-vacancy centers in a diamond lattice. The projections 
of an external magnetic field 
along the four nitrogen-vacancy axes are also shown. (b) The energy-level diagram of a nitrogen-vacancy center.}
\end{figure}

A negatively-charged nitrogen-vacancy (NV) center in diamond consists of a nitrogen atom adjacent to a vacancy in the carbon lattice (see Fig.~\ref{fig_nv_centers}). The hybridization of the two unpaired electrons leads to a quantum system with the energy-level diagram shown in Fig.~\ref{fig_nv_energy}. The spin magnetic triplet is formed at the ground state ($^3$A), consisting of a sub-level $\ket{m_s=0}$ at its lowest energy and another two zero-field degenerate sub-levels $\ket{m_s=\pm 1}$ raised by $\sim$2.87~GHz. 
When an external magnetic field $\vv{B_{ext}}$ with a component $\vv{B_z}$ along the N-V axis (see Fig.~\ref{fig_nv_centers}) is applied, the $\ket{m_s=\pm 1}$ sub-levels are split apart (i.e. Zeeman effect). 
The photon frequency $\Delta f$ associated with such an energy gap is proportional to $|\vv{B_z}|$:
\begin{equation}
    \Delta f=f_+-f_-=2\gamma_e|\vv{B_z}|,
\end{equation}
where $\gamma_e$ is the gyromagnetic ratio and equals to 28~GHz/T, and $f_+$ and $f_-$ are the frequencies for the transitions from $\ket{m_s=0}$ to $\ket{m_s=+1}$ and $\ket{m_s=-1}$, respectively. We use $\Delta f$ to derive $\vv{B_z}$. 

\begin{figure} [!t]
\centering
\includegraphics[width=2.6in]{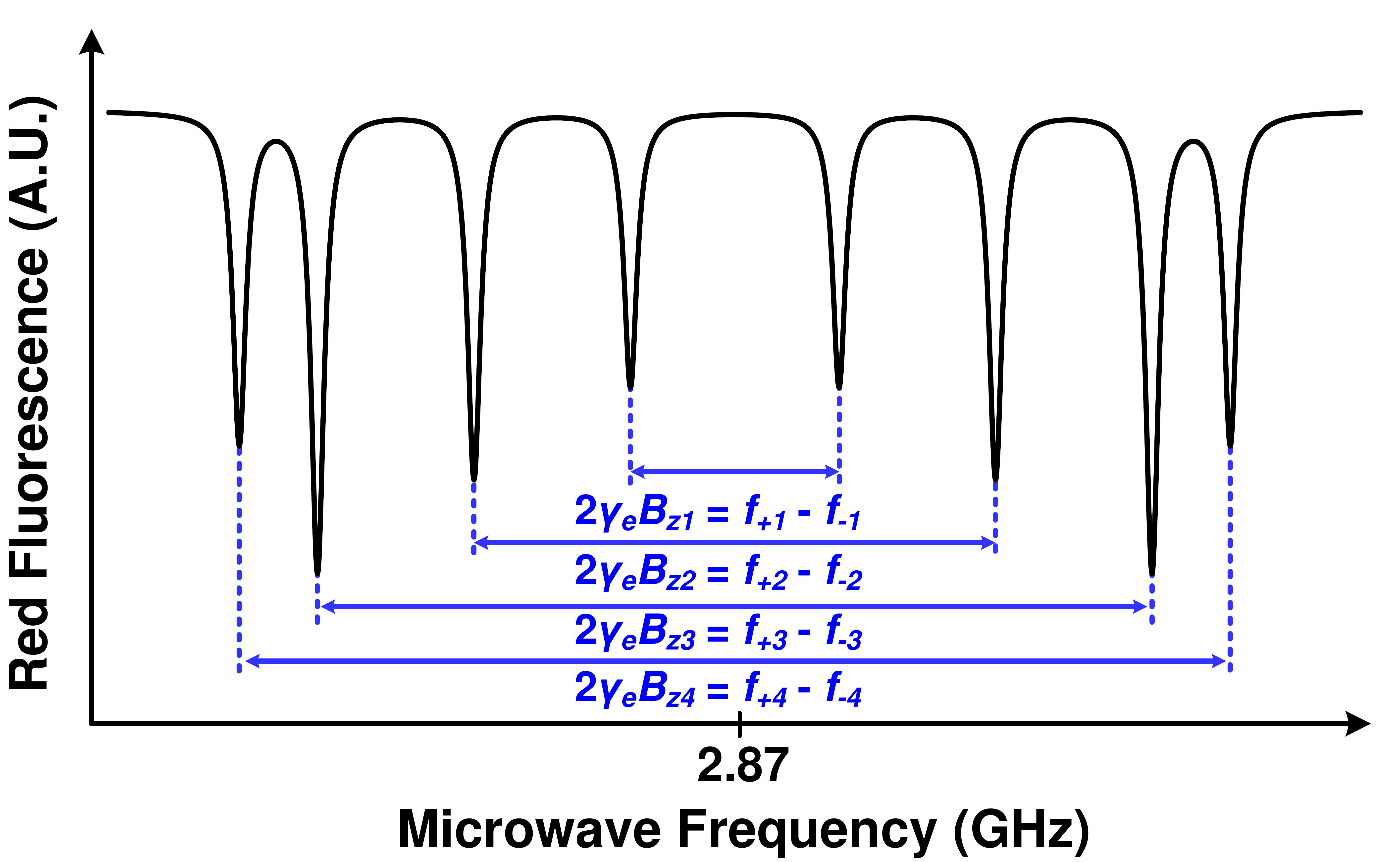}
\caption{The red-fluorescence intensity of the diamond at varying microwave frequency. An external magnetic-field bias with projections along the four N-V axes is assumed.}\label{fig_zeeman}
\end{figure}

NV magnetometry is performed by determining $f_+$ and $f_-$ via optically-detected magnetic resonance (ODMR)\cite{Gruber1997ScanningCenters}. In an ODMR experiment, the NV center spins are stimulated to their excitation states ($^3$E in Fig.~\ref{fig_nv_energy}) with green light ($\lambda\approx$~532~nm), and then relax back to the ground state ($^3$A). The relaxation of the $\ket{m_s=0}$ state is accompanied by bright red fluorescence ($\lambda\approx~$600$\sim$800~nm). In contrast, when the $\ket{m_s=\pm 1}$ states are excited and relax back, they can undergo an intersystem crossing into a metastable spin-singlet state ($^1$A in Fig.~\ref{fig_nv_energy}), and then into the $\ket{m_s=0}$ ground level\cite{Kim2019a} reducing the red fluorescence intensity. Thus, static or slowly-varying magnetic fields $\vv{B_{z}}$, can be determined by sweeping a microwave frequency $f_{RF}$ around 2.87~GHz and monitoring the average intensity. 
The observed resonances of Fig.~\ref{fig_zeeman} are $f_+$ and $f_-$, which give $\vv{B_z}$.

NV centers have four known orientations (Fig.~\ref{fig_nv_centers}) which separately lay along the tetrahedral axes of the host diamond. Accordingly, an external magnetic field $\vv{B_{ext}}$ has four projections -- $\vv{B_{z1}}$, $\vv{B_{z2}}$, $\vv{B_{z3}}$, $\vv{B_{z4}}$ -- along the NV orientations. This leads to four pairs of splitting in a single ODMR measurement (Fig.~\ref{fig_zeeman}). The magnetometer based on this principle, therefore, has a vector-field measurement capability by monitoring the different magnetic field projects and reconstructing $\vv{B_{ext}}$. That is advantageous over conventional Hall and fluxgate-based sensors, where three devices in \emph{x- y- z-} axes are needed for vector detection.

\begin{figure*}[!t] 
\centering
\includegraphics[width=6.5in]{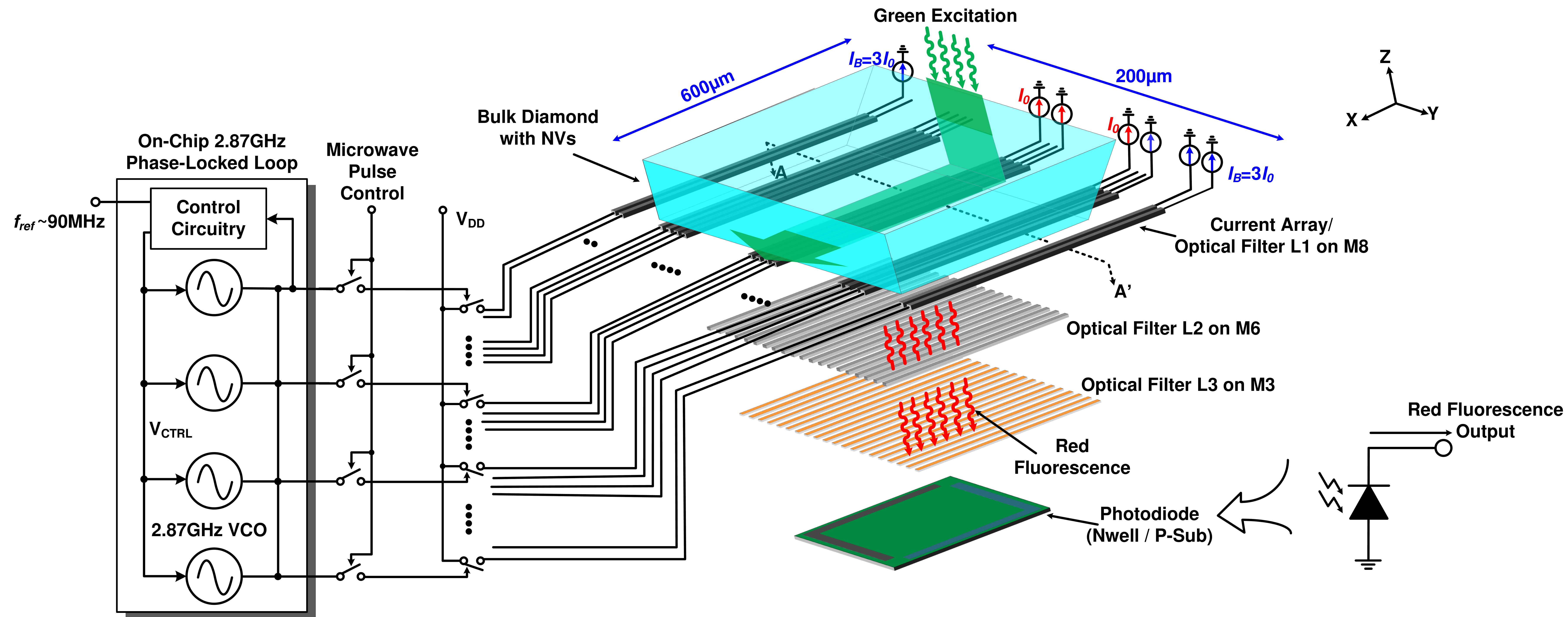}
\caption{The overall schematic of the CMOS quantum magnetometer with high scalability.}\label{fig_architecture}
\end{figure*}



To explore the feasibility of a chip-scale, low-cost quantum magnetometer, a custom-designed CMOS prototype was realized and reported, for the first time, in \cite{Ibrahim2018}. This chip, using TSMC 65-nm CMOS technology, integrates most of the critical components (except the green light source) for the ODMR operation. Using this hybrid CMOS-NV-center integration platform, the ODMR spectrum of a nanodiamond layer attached on top of the chip is demonstrated. The estimated sensitivity of the system is 74~$\upmu$T/Hz$^{1/2}$. Since the lattice orientations in the nanodiamond particles are random, the amount of frequency splitting in each NV is also random. As a result, this prototype cannot be used for vector-field sensing. Recently, we attached a film of bulk diamond (with uniform and well-defined lattice structure) to the same CMOS chip and demonstrated vector-sensing capability with 32~$\upmu$T/Hz$^{1/2}$ sensitivity\cite{Kim2019a}. 

Our first CMOS prototype demonstrates the basic concept of chip-scale miniaturization of NV-center quantum sensors. The achieved sensitivity is still limited by two main factors. (i) In our experiments we observed that the dominant noise source was the shot noise due to the green light despite the presence of the grating filter \cite{Kim2019a}. Ideally, this sensor would be limited to the red fluorescence shot noise and therefore, a higher green-to-red suppression ratio is required for the integrated photonic filter. We estimated that, for the diamonds used in \cite{Ibrahim2018,Kim2019a}, the intensity of red fluorescence is about 40$\sim$50~dB lower than that of the incident green light. Ultimately, an additional 30~dB out-of-band rejection is required for the photonic filter, so that the majority of photodiode noise is no longer generated by the green background. (ii) The sensing area used is only 50~$\upmu$m$~\times~$50~$\upmu$m. This limits the number of NV centers, $N$. Increasing $N$ improves the sensitivity due to larger red-fluorescence intensity (hence the \emph{SNR} of the ODMR spectrum). This can be achieved by using a larger sensing area and higher NV-center density in diamond. Note that the \emph{SNR} has the following dependency with $N$:
\begin{equation}
    \text{\it SNR}~\text{is} \begin{cases}
\propto N &\text{when noise is green-light limited}\\
\propto \sqrt{N} &\text{when noise is red-light limited}
\end{cases}.
\end{equation}

\section{A Scalable CMOS-NV Magnetometer for Enhanced Sensitivity}\label{sec_arch}

In this section, the design details of a new-generation CMOS-NV magnetometer\cite{Ibrahim2019ACapability} are provided. The highly scalable architecture of our CMOS-NV magnetometer allows for the microwave driving of NVs over a large area. The photodiode noise is further reduced with the adoption of a new on-chip photonic filter. These structures are co-designed with the on-chip electronics. 

\subsection{Systematic Architecture of the Chip} 

The overall schematic of the CMOS-NV magnetometer is given in Fig.~\ref{fig_architecture}. Shown on the left of Fig.~\ref{fig_architecture} is an on-chip phase-locked loop (PLL) which generates the 2.87-GHz microwave signal (see Section~\ref{sec_PLL}). To drive the NVs with the on-chip generated  microwave field, an array of current-driven linear wires are implemented using the M8 of the chip, of which the driving currents are toggled by the PLL output. Such a design addresses a major challenge regarding the uniformity of the microwave magnetic field over a large area. In Section~\ref{sec_launcher}, detailed explanations of the microwave launcher are provided. In this work, a diamond area of $\sim$500$~\times$~500~$\upmu$m$^2$ is excited by the microwave. The sensing area with uniform microwave excitation is limted to 300~$\times$~80~$\upmu$m$^2$ as discussed in Section~\ref{sec_co_design}. Shown in Section~\ref{sec_filter} and \ref{sec_co_design}, the current-driven wire array, along with additional two layers of metal gratings, also form a photonic filter in order to suppress the green light transmitted through the diamond placed on top of the chip. Finally, the spin-dependent red fluorescence of the NV centers is measured using a n-well/p-substrate photodiode.

\subsection{Generation of High-Homogeneity Magnetic Field}\label{sec_launcher}

Our NV-CMOS sensor interrogates an ensemble of NV centers to perform magnetometry. The microwave field strength determines both the ODMR resonance amplitude (contrast) and the resonance linewidth. Optimizing the sensitivity requires maximizing the contrast while minimizing the resonance linewidth \cite{Dreau2011AvoidingSensitivity}. For ensembles, the delivery of a homogeneous microwave magnetic field is critical in order to simultaneously perform this optimization across the entire area \cite{Eisenach_Loop}. Such microwave homogeneity is also critical to pulse-based coherent quantum control protocols, such as Ramsey-type sequence, which can  significantly  increase  the  sensitivity  to time-varying  external  magnetic  field \cite{Taylor2008High-SensitivityResolution}. Homogeneous microwave \emph{synchronously} rotate the spin states of a large number of NV centers on the Bloch sphere\cite{Bayat2014,Zhang2016}. Since the microwave-field strength determines the Rabi nutation frequency of each NV electron spin, spatial variation of the microwave field causes dephasing of the overall quantum ensemble.


Traditional microwave-launching structures include single straight wires\cite{Dreau2011AvoidingSensitivity,LeSage2013OpticalCells}, metal loops\cite{Stanwix2010CoherenceDiamond,Clevenson2014,Steinert2013MagneticResolution} and split-ring resonators\cite{Bayat2014,Zhang2016}. They can only keep the field homogeneity in an area that is much smaller than the launcher size. That is undesired for compact chip implementation when excitation of a large-size diamond is pursued. Meanwhile, the above structures also have poor power-delivery efficiency, hence watt-level microwave input power is common\cite{Clevenson2014,Bayat2014,Zhang2016}. We note the above solutions share one commonality in that they all rely on \emph{one piece of passive structure driven by a single electrical port}. That, unfortunately, makes it extremely difficult, if not impossible, to synthesize a certain desired current distribution\footnote{For electrically-small structures (size$\ll$$\lambda_{\text{2.87GHz}}$), the distributions of the current on the structure and the generated near-field magnetic wave follow a one-to-one mapping. Note that although displacement current also generates magnetic field, our following discussions are constrained to structures mainly with conduction current.}, because the only design variable is the structural geometry.

\begin{figure}
\centering
\includegraphics[width=3.3in]{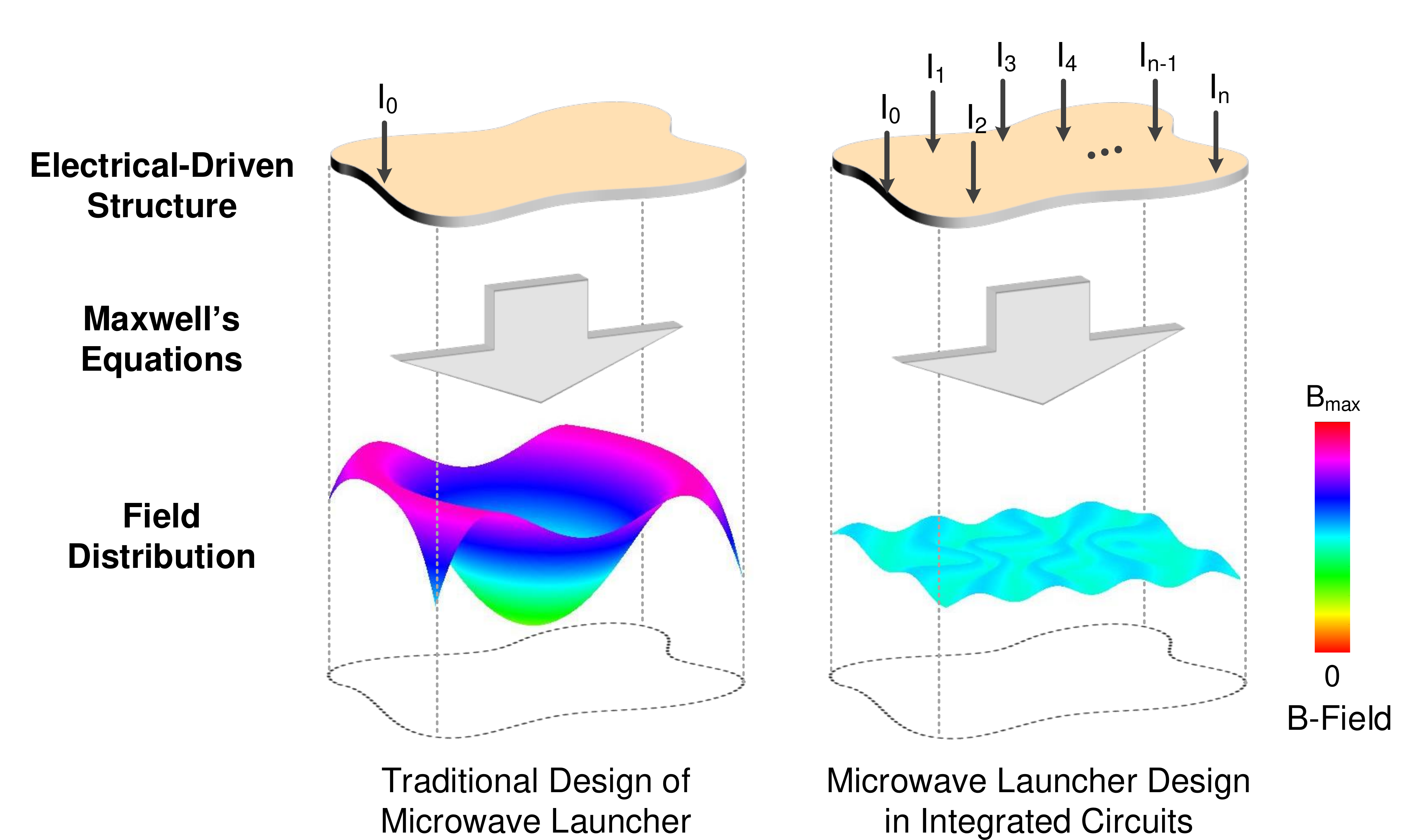}
\caption{Comparison of design methodologies for the microwave launcher in a quantum sensor. A distributed co-design of passive and active components is equivalent to adding more boundary conditions to an electromagnetic-solving problem, hence providing better control of the near-field pattern.}\label{fig_field_distribution}
\end{figure}

One distinct advantage of the implementation using integrated circuits is \emph{the highly flexible and tight integration between passive and active components.} A large number of biased transistors, when forming current sources, can be used to mandate the complex current values at various particular locations on the passive structure (Fig.~\ref{fig_field_distribution}). Compared to the aforementioned passive-only structures, this new design methodology provides abundant additional degrees of freedom through the positions and currents of the embedded transistors. A finer control of near-field wave distribution is therefore much easier. Note that similar concepts were already proposed and applied to millimeter-wave and terahertz integrated circuit designs\cite{Hajimiri2009,Sengupta2012ARadiators,Hu2018}.

\begin{figure}
\centering
\subfloat[]{\includegraphics[width=3in]{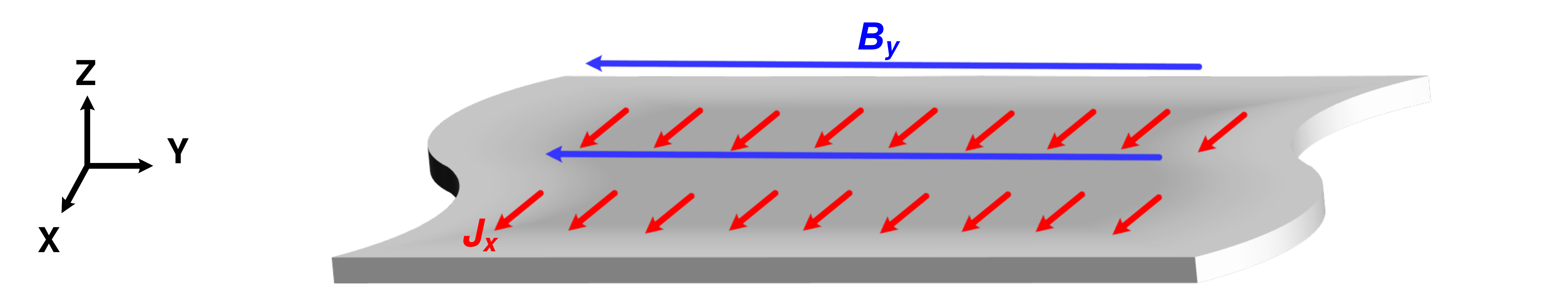}\label{fig_infinite_current_sheet}}\\
\subfloat[]{\includegraphics[width=3in]{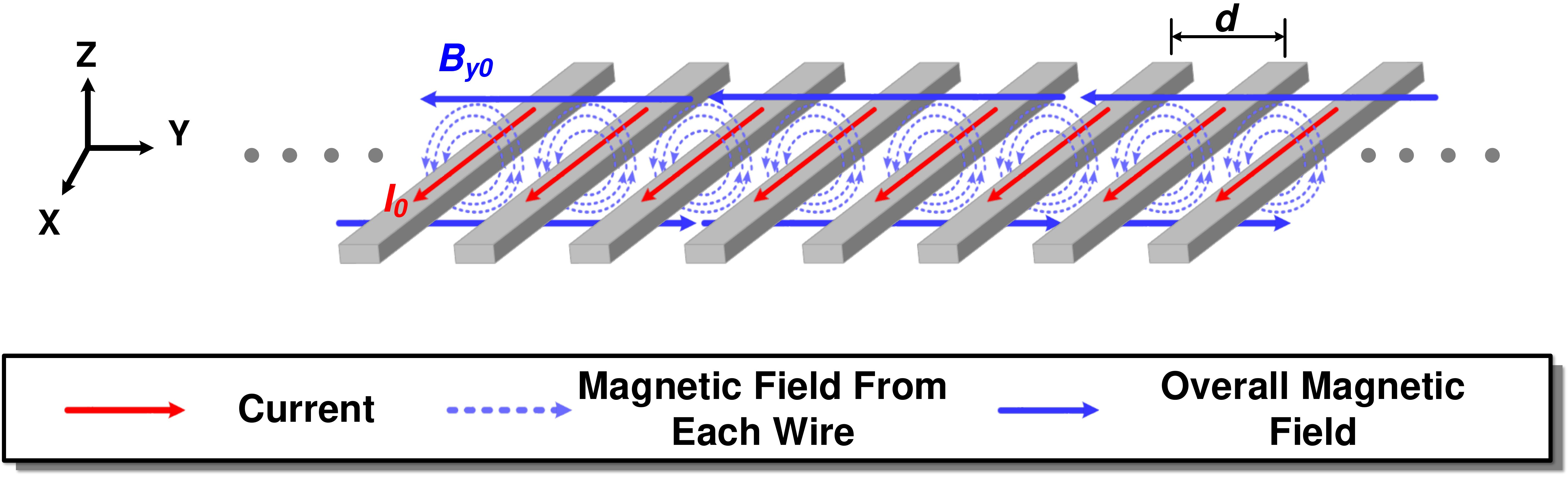}\label{fig_infinite_wire_array}}
\caption{Approaches to generate homogeneous magnetic field: (a) an infinite sheet of uniformly-distributed current, (b) an infinite array of wires with uniform driving current.}
\end{figure}

Following the above methodology, our microwave launcher design in Fig.~\ref{fig_architecture} evolves from an ideal, infinite sheet of uniformly-distributed surface current density $J_x$ (see Fig.~\ref{fig_infinite_current_sheet}), which generates a homogeneous field $B_y$ with a transverse ($y$-) direction and strength of $\mu_0J_x/2$ ($\mu_0$: permeability of vacuum). However, such current uniformity is hard to imitate in a single metal structure, due to the self-redistribution of current like skin effect. To prevent such redistribution, the current sheet is first transformed into an infinite array of wires, each driven by an identical current $I_0$ (Fig.~\ref{fig_infinite_wire_array}). With a tight wire center-to-center pitch $d$ (or equivalent current density of $I_0/d$), a uniform transverse magnetic field can still be obtained, with a field strength of $\mu_0I_0/2d$. The wire array also ensures high scalability in the longitudinal ($x$-) direction for large-area diamond. 

\begin{figure}[t]
\centering
\includegraphics[width=3in]{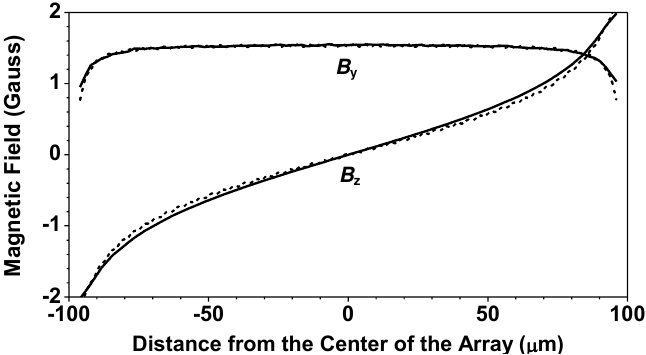}
\caption{Simulated and calculated field distribution ($f$=2.87~GHz) of a wire array with uniform driving current ($I_0$=1~mA).}\label{fig_sim_finite_array_uniform}
\end{figure}

\begin{figure}
\centering
\subfloat[]{\includegraphics[width=3.5in]{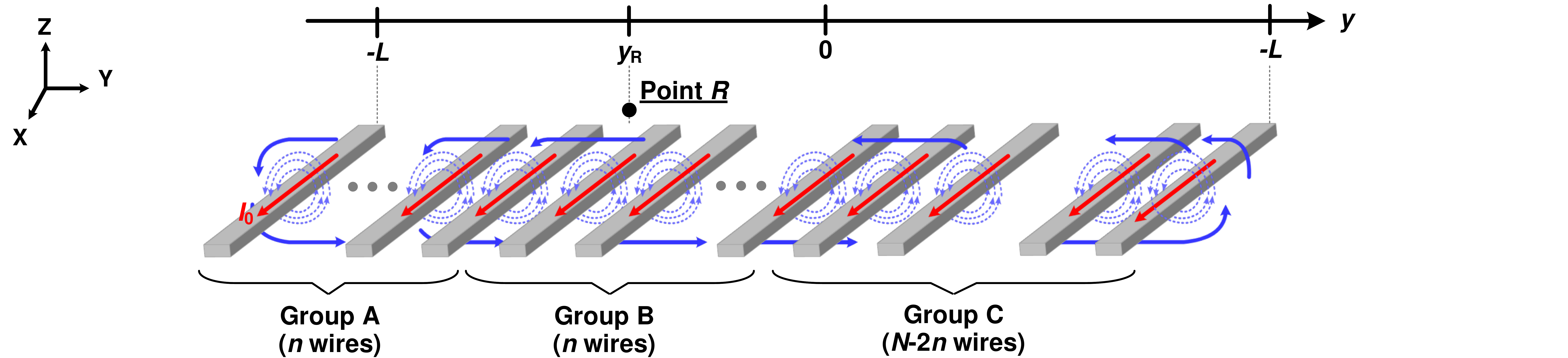}\label{fig_finite_wire_array}}\\
\subfloat[]{\includegraphics[width=3.5in]{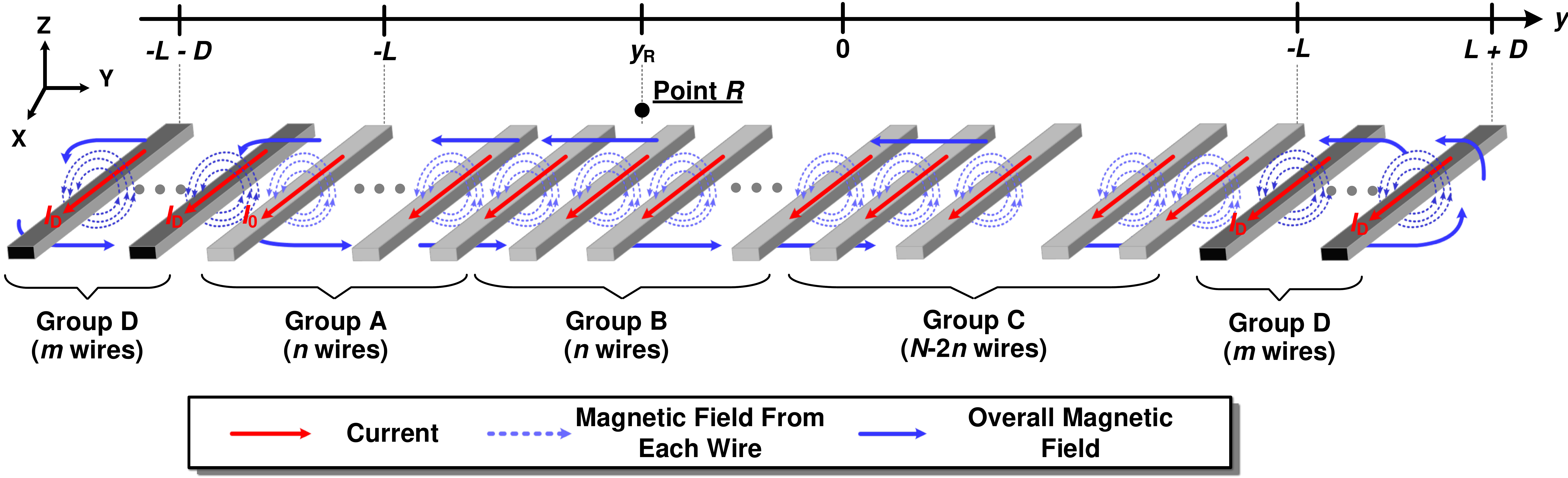}\label{fig_finite_wire_array_w_boundary}}
\caption{(a) A finite array of wires with uniform driving current. (b) An additional pair of boundary wire arrays (Group~\emph{D}) for nulling the vertical magnetic field between $-L$ and $+L$.}
\end{figure}

In reality, only a finite wire array can be implemented. The introduced boundaries, however, break the above homogeneity. Fig.~\ref{fig_sim_finite_array_uniform} shows the simulated field distribution of a 48-wire array with $d$=4~$\upmu$m. Although the transverse component of the field $B_y$ maintains homogeneity, the vertical component $B_z$ becomes non-zero and exhibits large gradient over $y$-axis. To understand that, we focus at a reference point (Point~\emph{R}) located above an array of $N$ wires and closer to the left side (i.e. $y_{R}<0$ in Fig.~\ref{fig_finite_wire_array}). The wires are then divided to three groups: the \emph{n} wires located between $y_R$ and the left boundary $-L$ (Group~\emph{A}), another \emph{n} wires symmetrically located at the right side of Point~\emph{R} (between $y_R$ and 2$y_R+L$, Group~\emph{B}), and the rest (Group~\emph{C}). Next, note that at Point~\emph{R}, the vertical magnetic fields generated by Group~\emph{A} and \emph{B} cancel, and the $B_z$ generated by Group~$C$, by Ampere's Law, is:
\begin{equation}\label{eqn_array_field}
  B_{z,C}(y_R)=-\frac{1}{2\pi}\int_{2y_R+L}^{+L}\frac{\mu_0J_xdy}{y-y_R}=\frac{\mu_0J_x}{2\pi}\ln\frac{L+y_R}{L-y_R},
\end{equation}
where the discreteness of the array is approximated as a uniform current sheet with a current density of $J_x=I_0/d$. Note that (\ref{eqn_array_field}) also applies for $y_R>0$. Shown in Fig.~\ref{fig_sim_finite_array_uniform}, the result calculated from (\ref{eqn_array_field}) well matches the simulation by HFSS\cite{ANSYSHFSS}. Intuitively, $B_z$ increases towards the edges because Group~\emph{C} has a larger total current there and gets closer to Point~\emph{R}.

\begin{figure}[t]
\centering
\includegraphics[width=3in]{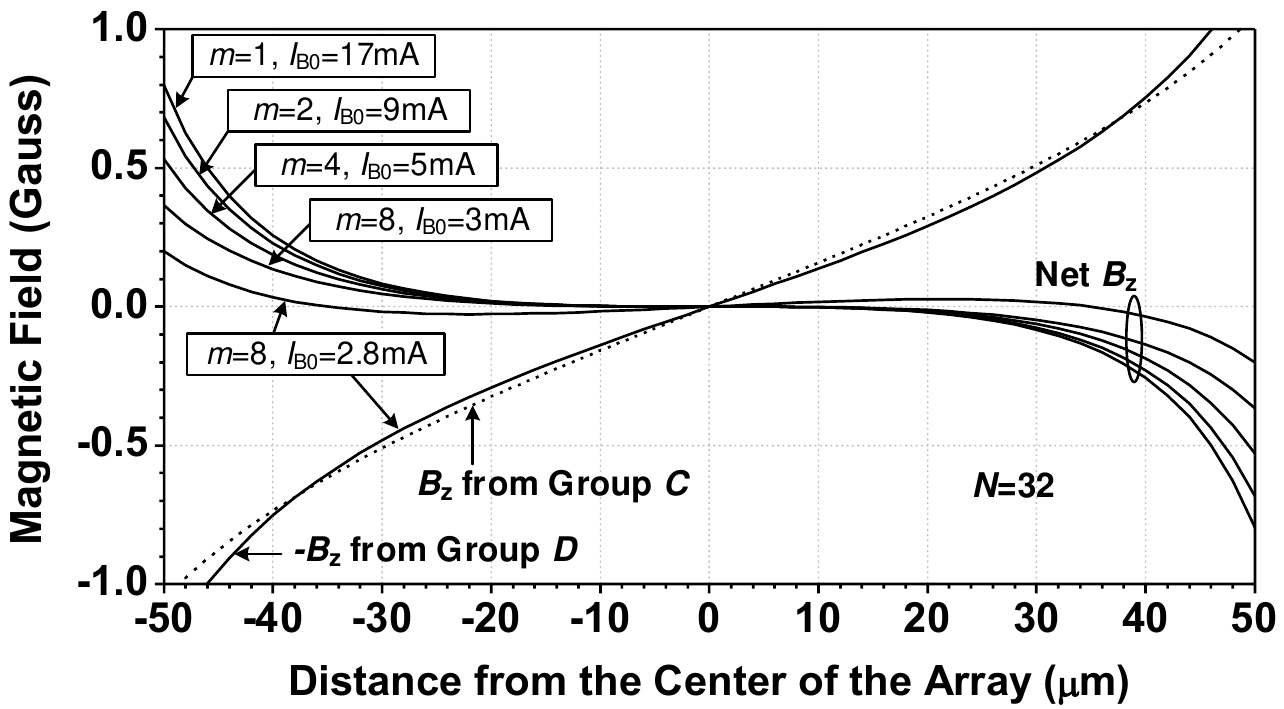}
\caption{Calculated profiles of vertical field generated by Group~\emph{C} and \emph{D} (and the residue after the cancellation) of the array shown in Fig.~\ref{fig_finite_wire_array_w_boundary}. Note that the $B_z$ of Group~\emph{D} is plotted with its polarity reversed, to facilitate a straightforward comparison with $B_z$ from Group~\emph{C}.}
\label{fig_cal_finite_boundary}
\end{figure}

\begin{figure}
\centering
\includegraphics[width=3in]{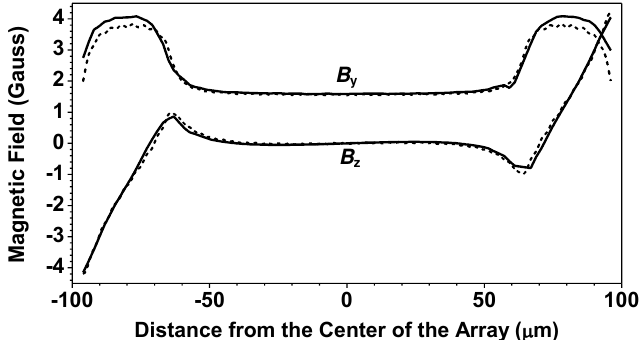}
\caption{Simulated (solid line) and calculated (dashed line) field distribution of the entire launcher wire array. Here, $f$=2.87~GHz, $N$=32, $m$=8, $I_0$=1~mA, $I_D$$\approx$3~mA.}
\label{fig_sim_finite_array_final}
\end{figure}

To create an opposite magnetic-field gradient for nulling the above $B_z$, a pair of additional $m$-wire arrays  (Group~\emph{D} in Fig.~\ref{fig_finite_wire_array_w_boundary}), which are driven by $I_D$ per wire, are symmetrically placed at $y=-L-D$ to $-L$ and $y=L$ to $L+D$. Their combined impact on the vertical magnetic field at $y=y_R$ is derived as:
\begin{align}\label{eqn_array_boundary}
 B_{z,D}(y_R) =&\int^{-L}_{-L-D}\frac{\mu_0 mI_Ddy}{2\pi D(y_R-y)}-\int^{L+D}_{L}\frac{\mu_0 mI_Ddy}{2\pi D(y-y_R)}\nonumber\\
 =&\frac{m\mu_0I_D}{2\pi D}\ln\frac{(y_R+L+D)(L-y_R)}{(L+D-y_R)(y_R+L)},
\end{align}
of which some example plots (to be discussed more) is shown in Fig.~\ref{fig_cal_finite_boundary}, indicating the similarity of its curve shape to that from Group~\emph{C}. To further determine the value of $I_D$, we compare the derivatives of (\ref{eqn_array_field}) and (\ref{eqn_array_boundary}) around the center of the launcher:
\begin{align}
    \frac{dB_z}{dy}|_{y=0}= \begin{cases}
\frac{\mu_0J_x}{\pi L} &\hspace{1cm}\textrm{for Group~\emph{C}}\\
-\frac{\mu_0 mI_D}{\pi L(L+D)} &\hspace{1cm}\textrm{for Group~\emph{D}}
\end{cases}.
\end{align}
Therefore, for zero $B_z$ around the array center, the value of $I_D$ should be:
\begin{equation}\label{eqn_array_condition}
    I_{D,0}=\frac{L+D}{m}J_x=\frac{L+D}{md}I_0=(1+\frac{N}{2m})I_0.
\end{equation}
The last step of (\ref{eqn_array_condition}) assumes that the boundary arrays adopt the same wire pitch $d$ as the $N$-wire uniform array in the center
For a 32-wire uniform array ($N$=32) with $I_0$=1~mA and $d$=4~$\upmu$m, the vertical field $B_{z,D}$ generated by the boundary array pair with varying wire numbers $m$, as well as the residual of $B_z$ after cancellation, are plotted in Fig.~\ref{fig_cal_finite_boundary}. We see that a larger $m$ improves the area of uniformity at a modest expense of higher current consumption. Interestingly, if the $I_D$ value is slightly lower than the one calculated by (\ref{eqn_array_condition}) (e.g. for $m$=8, the calculated value for $I_D$ is 3~mA), the $B_z$ field around $y$=0 is not cancelled perfectly but the overall 95\% uniformity area is extended to $y$$\approx$$\pm$40~$\upmu$m. Lastly, Fig.~\ref{fig_sim_finite_array_final} shows the HFSS-simulated field distribution of the entire launcher ($N$=32, $m$=8, $d$=4~$\upmu$m), which is implemented at the Metal 8 (M8) interconnect layer of our chip prototype. The calculated profile in Fig.~\ref{fig_sim_finite_array_final} is achieved with $I_D$=2.8$I_0$. Note that the HFSS simulation is done with a slightly larger $I_D$ ($I_D$=3$I_0$) to account for the uncalculated effect of current in the return path. Compared to Fig.~\ref{fig_sim_finite_array_uniform}, the overall homogeneity of the magnetic vector field is significantly improved.

\subsection{Nano-Photonic Filter on CMOS}\label{sec_filter}

Plasmonic filters based on a multi-stacked-layer gratings in the BEOL of CMOS technologies have been demonstrated previously in \cite{hong2017fully}. In our first CMOS quantum sensor\cite{Ibrahim2018}, a single-layer filter based on the same principle of wavelength-dependent plasmonic loss is adopted, which exhibits a green-to-red rejection ratio of $\sim$10~dB in our experiment. Based on the discussion in Section~\ref{sec_physics}, that ratio should be improved to enhance the magnetometry sensitivity. Next, we show how a wavelength-dependent diffraction pattern under the plasmonic filter layer is utilized to achieve the goal. To facilitate that discussions, we first revisit some details related to the plasmonic grating layer.

\begin{figure}[h]
\centering
\includegraphics[width=3in]{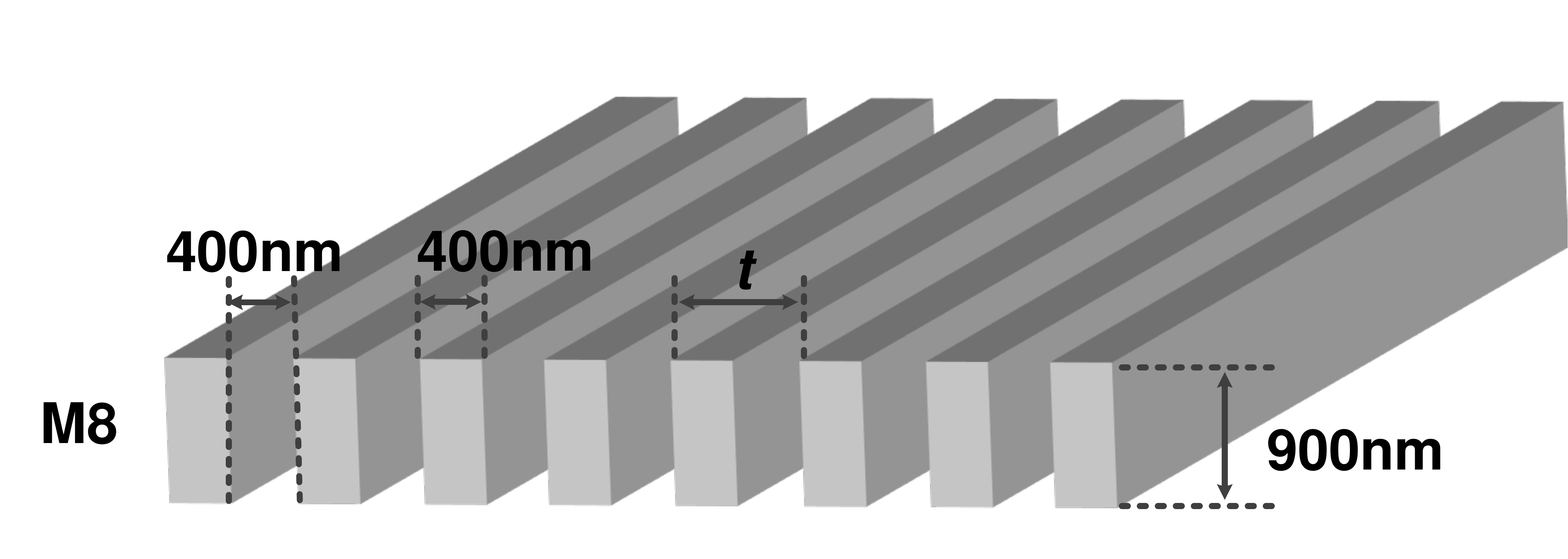}
\caption{A single-layer plasmonic grating filter implemented on Metal~8\cite{Ibrahim2018}\label{fig_single_layer_filter}.}
\end{figure}

\begin{figure}[t]
\centering
\subfloat[]{\includegraphics[width=2.4in]{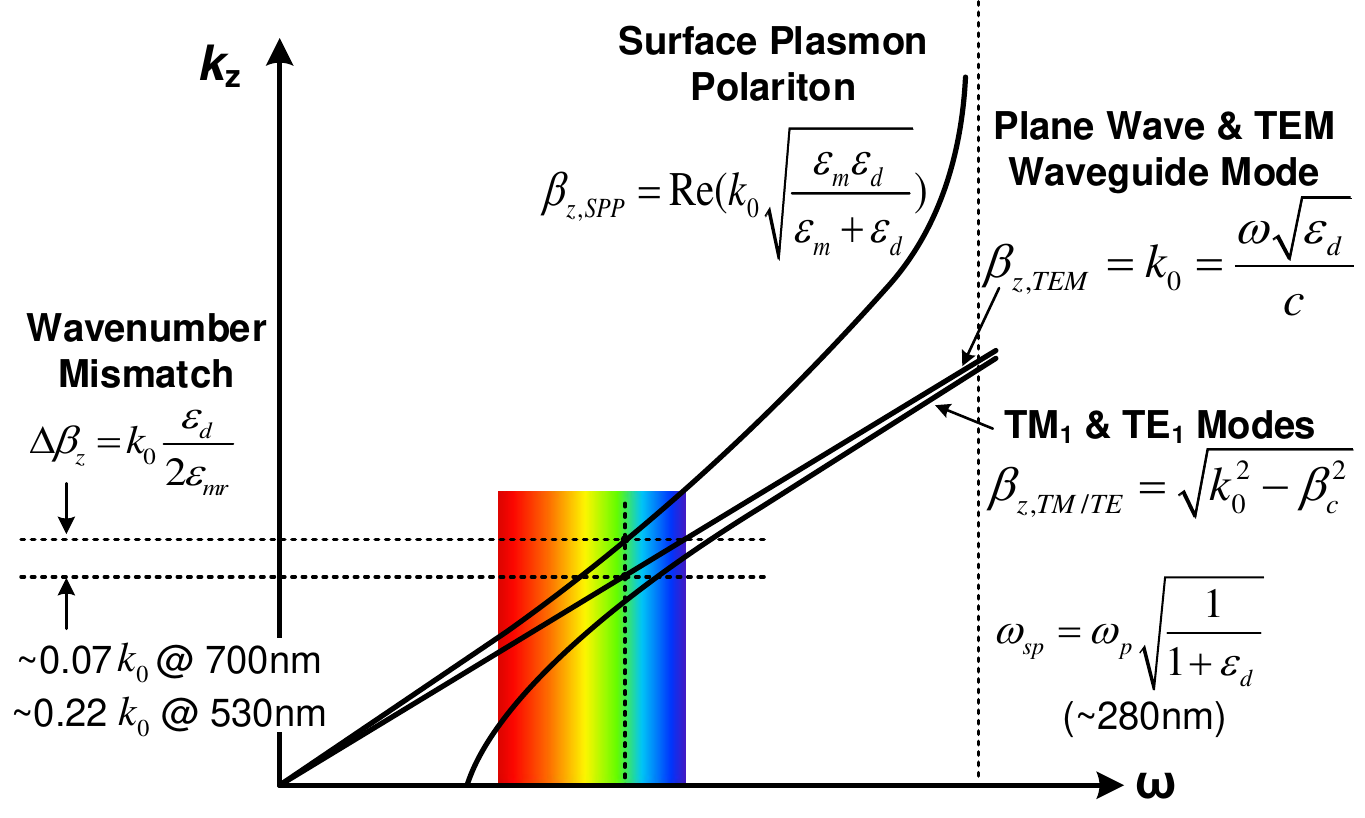}\label{fig_dispersion}}\hfill
\subfloat[]{\includegraphics[width=1in]{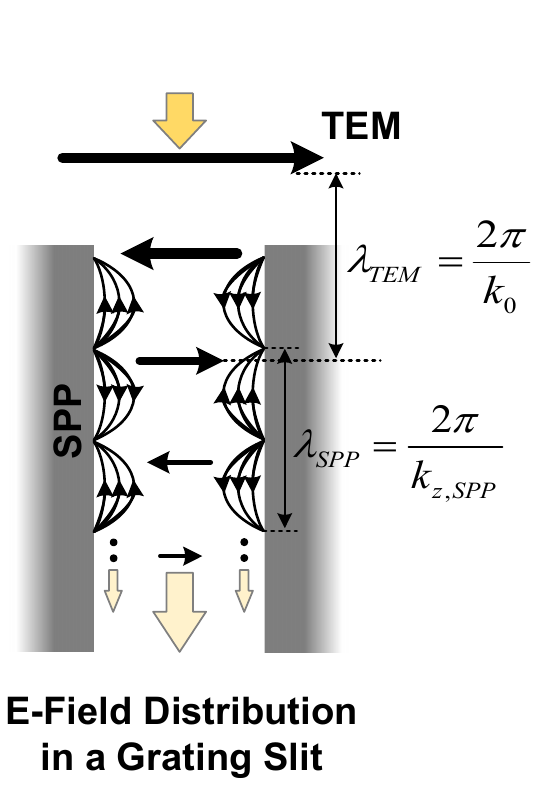}\label{fig_mode_coupling}}
\caption{TEM and SPP modes inside a grating slit (modeled as a parallel-metal-plate waveguide): (a) dispersion relationships and (b) short-distance coupling.}
\end{figure}

\begin{figure}[!t]
\centering
\subfloat[]{\includegraphics[width=3.1in]{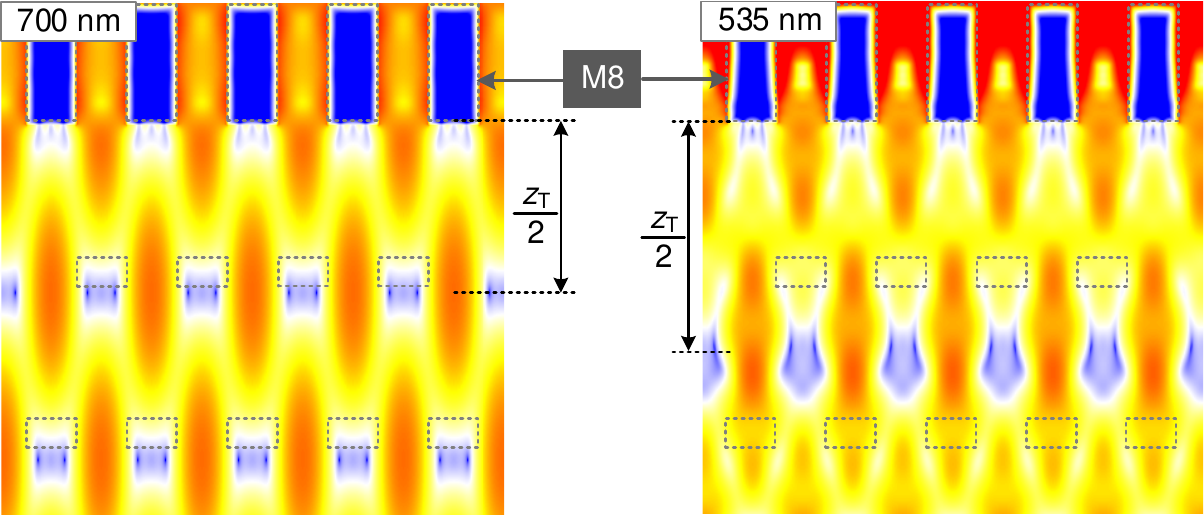}\label{fig_M8_grating_patterns}}\\
\subfloat[]{\includegraphics[width=3.1in]{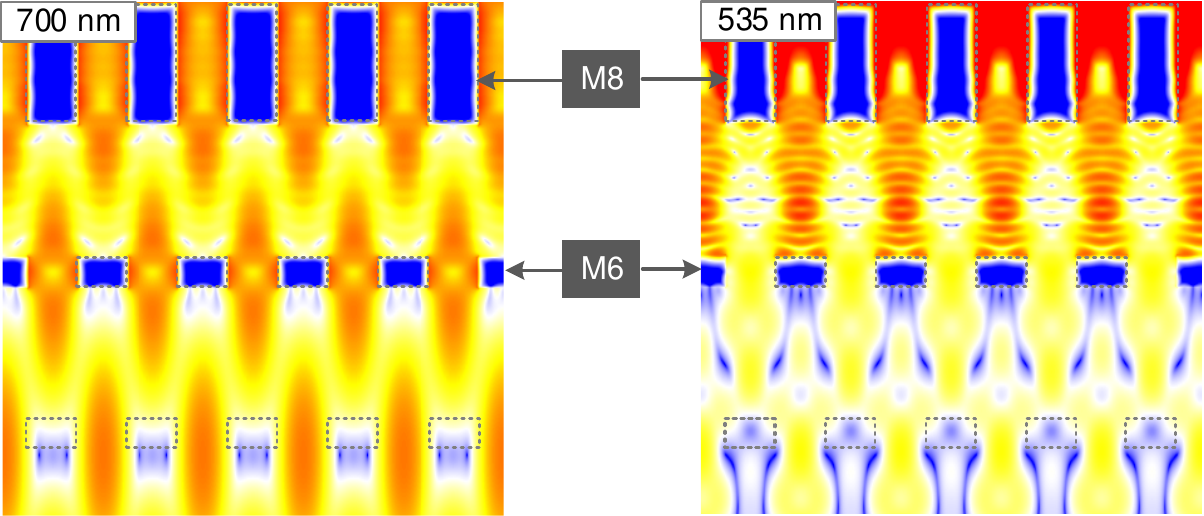}\label{fig_M8_M6_grating_patterns}}\\
\subfloat[]{\includegraphics[width=3.1in]{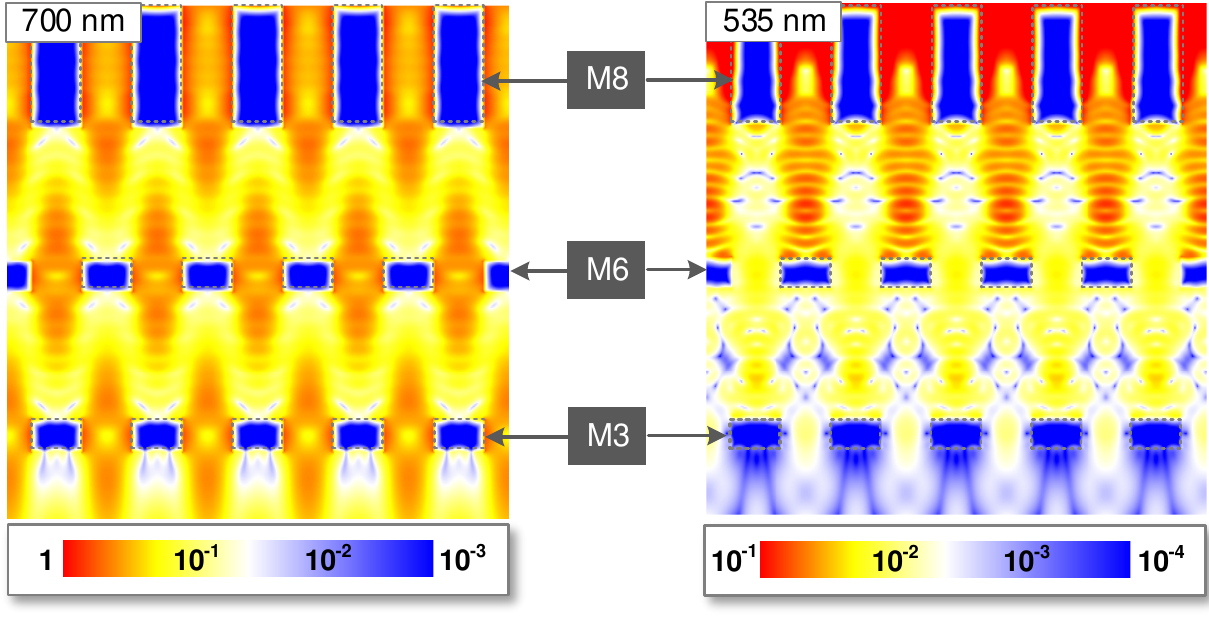}\label{fig_M8_M6_M3_grating_patterns}}\\
  \caption{The simulated FDTD Poynting vector profile $|P|$=$|E\times H|$ for (a) single-layer grating in M8, (b) double gratings in M8 and M6, and (c) triple-layer gratings in M8, M6 and M3. {\hspace{3mm} Note:} the scale bar for green (right column) is 10$\times$ lower than that for red (left column), in order to count for the $\sim$10-dB plasmonic loss in the M8 grating and to better show M6 and M3 gratings further suppress the transmission at 535~nm.}\label{fig_grating_patterns}
\end{figure}

Figure~\ref{fig_single_layer_filter} shows the structure of the filter used in \cite{Ibrahim2018,Kim2019a}. Each slit is considered as a parallel plate waveguide transmitting light inside a dielectric (refractive index $\sqrt{\epsilon_d}$$\approx$1.5) in the $z$- direction. The incoming light is modeled as a plane wave with a transverse electrical field $E_x$ and a propagation constant $k_0=\omega\sqrt{\epsilon_d}/c$ ($c$ is the speed of light in vacuum). This light is coupled to the TEM mode of the parallel-plate waveguides, which has identical propagation constant $k_0$ (see Fig.~\ref{fig_dispersion}). As the propagating wave interacts with the metal, the surface plasmon polariton (SPP) mode at the metal-dielectric interface is excited. Note that the dispersion relation of SPP mode is:
\begin{align}\label{eqn_spp_dispersion}
    k_{z,SPP} &=\beta_{SPP}+j\alpha_{SPP}=k_0\sqrt{\frac{\epsilon_m}{\epsilon_m+\epsilon_d}}.
\end{align}
In (\ref{eqn_spp_dispersion}), the permittivity ($\epsilon_m$=$\epsilon_{mr}+j\epsilon_{mi}$) of the metal (copper in our case) has a negative real part, which leads to $\beta_{SPP}>k_0$. Although coupling from the TEM waveguide mode and the SPP mode generally does not occur\footnote{Normally, dedicated material/geometrical configurations, such as the Otto and Kretschmann-Raether configurations, are needed to enable the excitation of the SPP mode.}, it still does in this case. That is because, by examining the value of $\beta_{SPP}$ in (\ref{eqn_spp_dispersion}):  
\begin{align}\label{eqn_spp_beta}
    \beta_{SPP}=\Re(k_0\sqrt{\frac{\epsilon_m}{\epsilon_m+\epsilon_d}})\approx k_0(1-\frac{\epsilon_d}{2\epsilon_{mr}}),
\end{align}
we see that since $\abs{\epsilon_{mr}}$ remains large, the relative mismatch between the $\beta_{SPP}$ and $\beta_{TEM}$=$k_0$, namely $\epsilon_d/2\epsilon_{mr}$ in (\ref{eqn_spp_beta}), are $\sim$0.07 for red (700~nm) and $\sim$0.22 for green (530~nm) (see Fig.~\ref{fig_dispersion}). These calculations are based on the Drude-Brendel-Bormann model \cite{Rakic1998OpticalDevices,Hagemann1975OpticalAl2O3}. Along the vertical propagation distance of the grating (i.e. the M8 grating thickness $d$=900~nm, or 2$\lambda_{red}$=2.5$\lambda_{green}$ in dielectric), the accumulated phase mismatch is:
\begin{align}\label{eqn_phase_mismatch}
    \Delta\phi=2\pi\frac{\epsilon_d}{2\epsilon_{mr}}\frac{d}{\lambda},
\end{align}
or $\sim$50$^o$ (red) and $\sim$180$^o$ (green), respectively. Therefore, for red light, constructive coupling from the TEM mode to SPP mode still occurs within the single-layer grating. For green light, although phase mismatch appears to be large, the coupling is still effective in the M8 slit due to the large SPP loss; and when the TEM wave reaches the bottom half of the slit, its power is already heavily depleted. To quantify the loss of the SPP mode, we derive the attenuation factor $\alpha_{SPP}$ in (\ref{eqn_spp_dispersion}):   
\begin{align}\label{eqn_spp_alpha}
    \alpha_{SPP}=\Im(k_0\sqrt{\frac{\epsilon_m}{\epsilon_m+\epsilon_d}})\approx k_0\frac{\epsilon_d\epsilon_{mi}}{2\epsilon_{mr}^2},
\end{align}
$\alpha_{SPP}$ exhibits large difference between red ($\approx$0.006$k_0$) and green ($\approx$0.18$k_0$) as calculated based also on the Drude-Brendel-Bormann model.


\begin{figure}[!b]
\centering
\includegraphics[width=3.2in]{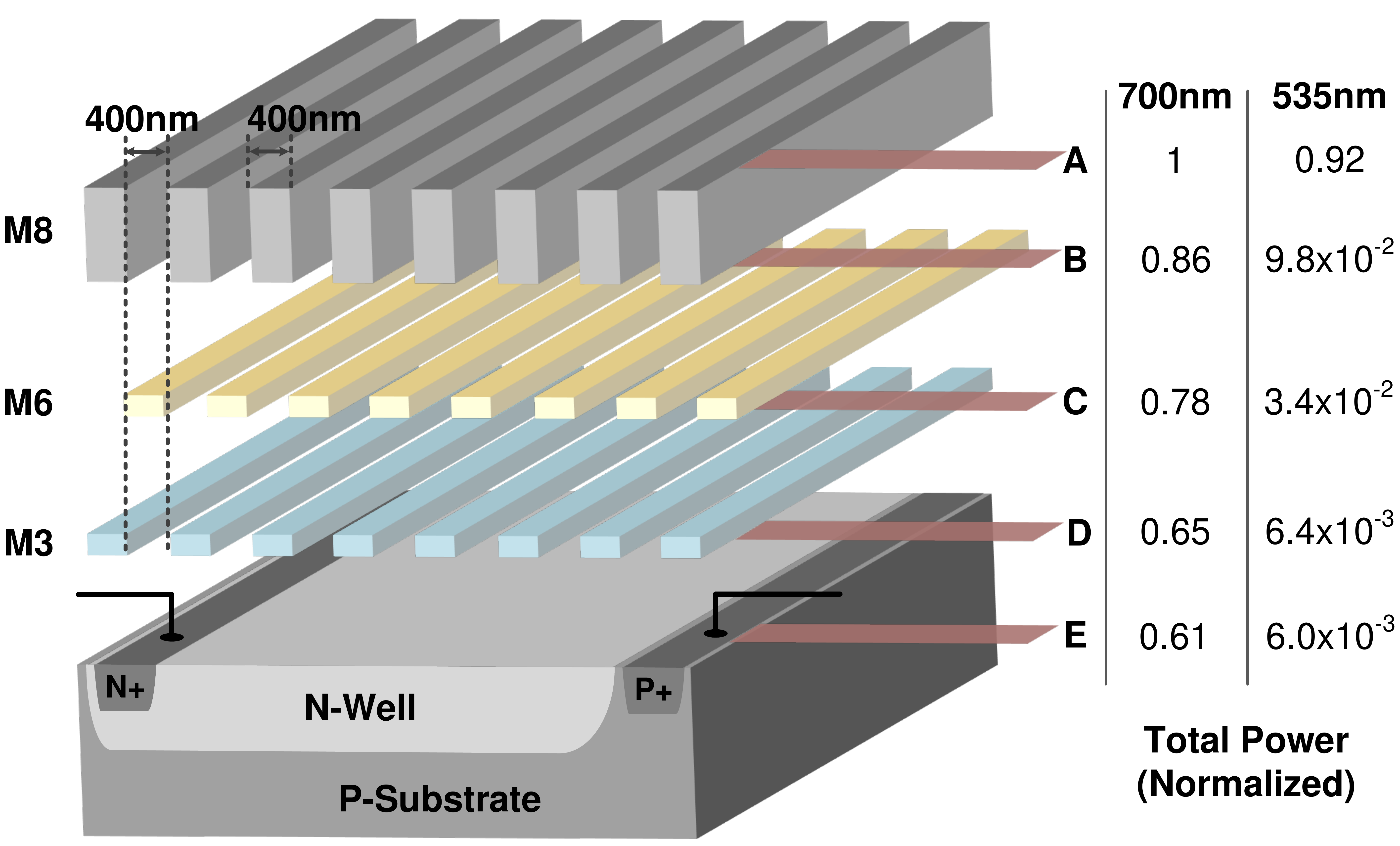}
\caption{Simulated transmission values through the filter. The total power is derived by surface integrals of Poynting vectors at varying depths of the chip.}\label{fig_three_layer_filter}
\end{figure}

\begin{figure}[!hb]
\centering
\includegraphics[width=3.2in]{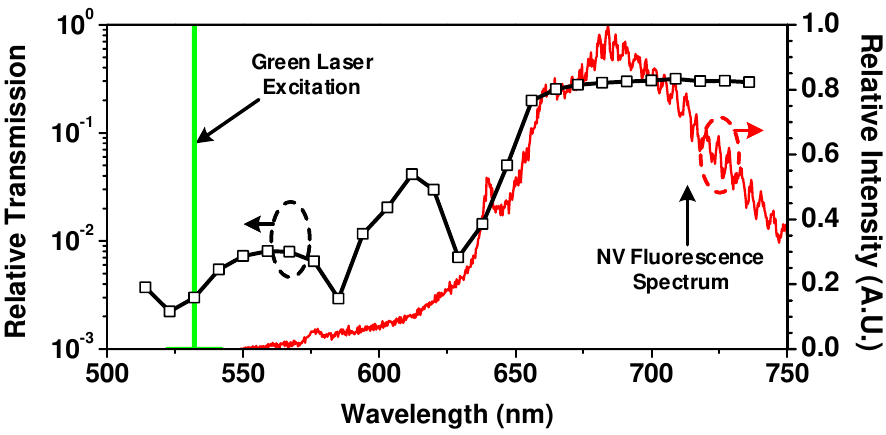}
\caption{The transmission through the filter at different wavelength.}\label{fig_filter_freq_response}
\end{figure}

\begin{figure*}[!b]
\centering
\includegraphics[width=6in]{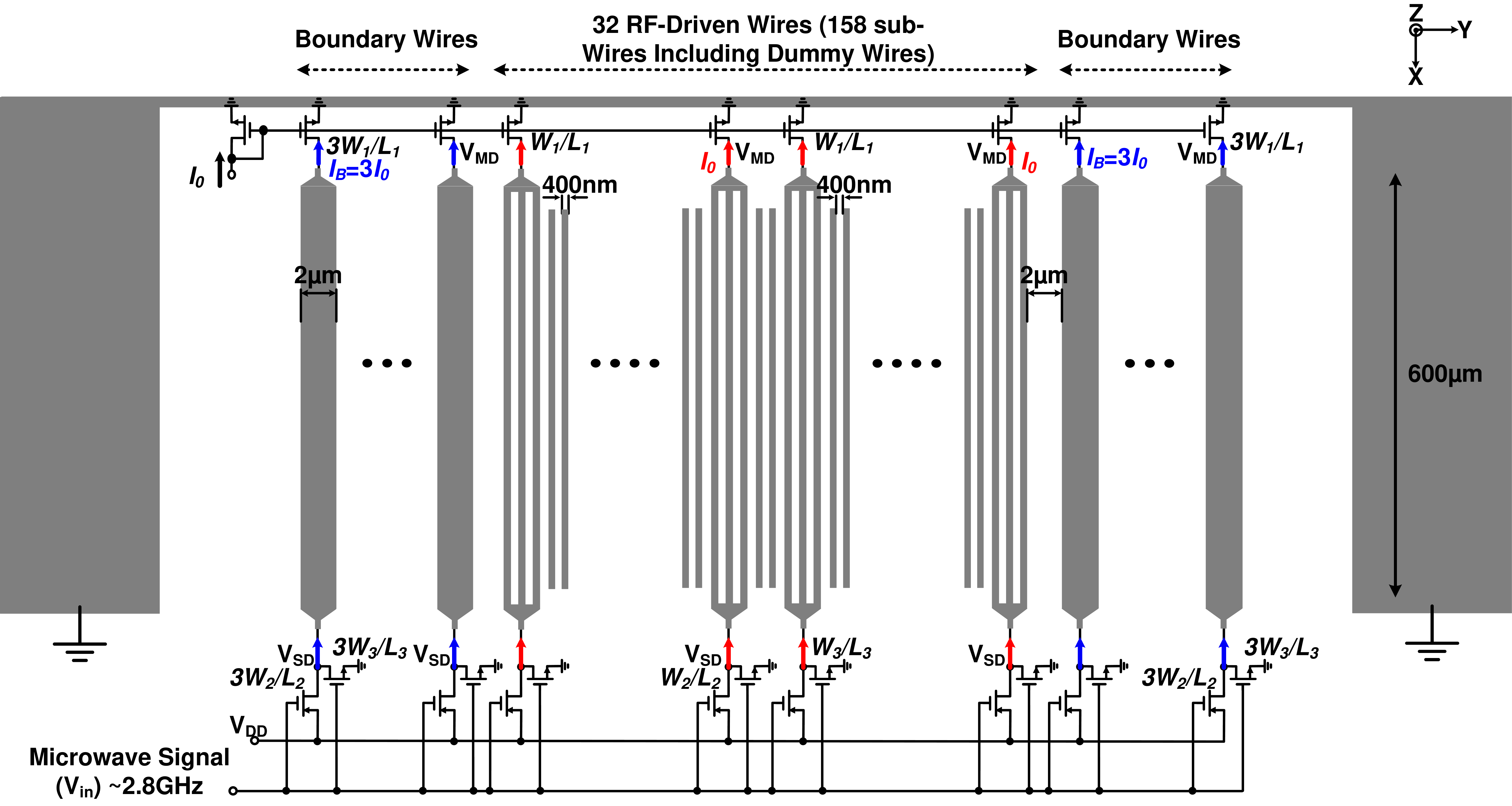}
\caption{Schematic of the microwave launcher with the switches and the current sources.}\label{fig_array_circuits}
\end{figure*}

Using a finite-difference time-domain (FDTD) solver, Lumerical, we simulate the green-to-red suppression ratio due to the slit in M8 layer ($d$=900~nm), which is $\sim$9~dB (Fig.~\ref{fig_M8_grating_patterns}). Increasing the slit thickness by stacking more grating layers (in M7, M6...) is expected to further increase the suppression of the green light\cite{hong2017fully}. However, that causes degradation of red-light transmission due to the scattering at the side walls formed by the sparse inter-layer via pillars. Moreover, as Section~\ref{sec_co_design} describes, the grating structure in M8 is also used for the microwave wire array (Section~\ref{sec_launcher}). Connecting it to the lower metal layers would significantly decrease the density and the uniformity of the RF current inside the wires. In our chip, an alternative approach based on the wavelength-dependent diffraction pattern of the grating is adopted.

As the simulation in Fig.~\ref{fig_M8_grating_patterns} shows, the diffraction of the light coming out of the grating in M8 causes repeated self-imaging patterns in the chip dielectric. That optical phenomenon is called Talbot effect\cite{Talbot1836FactsIV} and the vertical ($z$-) period of the self-images (i.e. Talbot length) is:
\begin{equation}\label{eqn_talbot_length}
    z_T=\frac{\lambda}{1-\sqrt{1-\frac{\lambda^2}{t^2}}},
\end{equation}
where $\lambda$ is the wavelength in the dielectric and $t$ is the grating period (800~nm in our case). For the red (700~nm) and green (535~nm) light, $z_T$ is $\sim$2.5~$\upmu$m and $\sim$3.4~$\upmu$m (see Fig.~\ref{fig_M8_grating_patterns}). That means we may strategically place additional metal structures at positions which are dark in red diffraction pattern (so that the light is not affected) but bright for green (so that the light is blocked). In the 65~nm CMOS process used for this work, the distance between M8 and M6 layers is close to the half Talbot length for red; therefore, a M6 grating with 400~nm strip width and pitch is placed. With the positions of the metal strips right under the slits of M8, the M6 grating has little interference with the red-light transmission, while blocking a significant portion of the green light (Fig.~\ref{fig_M8_M6_grating_patterns}). It is noteworthy that the similar multi-grating concept was previously adopted for lensless 3D CMOS cameras\cite{Wang2012}, where the angle sensitivity of the structure is utilized. To the authors' best knowledge, this is the first time in CMOS that the idea of diffraction selectivity is applied into optical spectral filtering.

Following the same principle, a third grating layer in M3, of which the distance to M8 is about one Talbot length at 700~nm, is added to further enhance the green-light rejection (Fig.~\ref{fig_M8_M6_M3_grating_patterns}). The same grating strip width and pitch of 400~nm are used. Fig.~\ref{fig_three_layer_filter} provides the normalized transmitted power (Poynting vector integrated over a $x$-$y$ cross-sectional area) at varying chip depths. For green light, the plasmonic behavior attenuates the power by 9.7~dB, and the M6 and M3 gratings pose an additional 12.1~dB loss, leading to a total attenuation of 21.8~dB. Meanwhile, the total insertion loss for red light is only 2.1~dB. Note that the reflection loss at the top of M8, which is $\sim$3~dB due to the 50\% metal fill factor, is not included in both cases.

Lastly, the simulated transmission response (excluding the 50\% surface reflection) of the filter from 500~nm to 750~nm is plotted in Fig.~\ref{fig_filter_freq_response}. The spectrum of the NV-center red fluorescence is also shown. It can be seen that the majority of the fluorescence power between 650 to 725~nm is transmitted efficiently to the photodiode beneath the filter.

\subsection{Co-Design of the Circuits, Microwave Launcher and Photonic Filter}\label{sec_co_design}

\begin{figure}[h]
\centering
\includegraphics[width=3.2in]{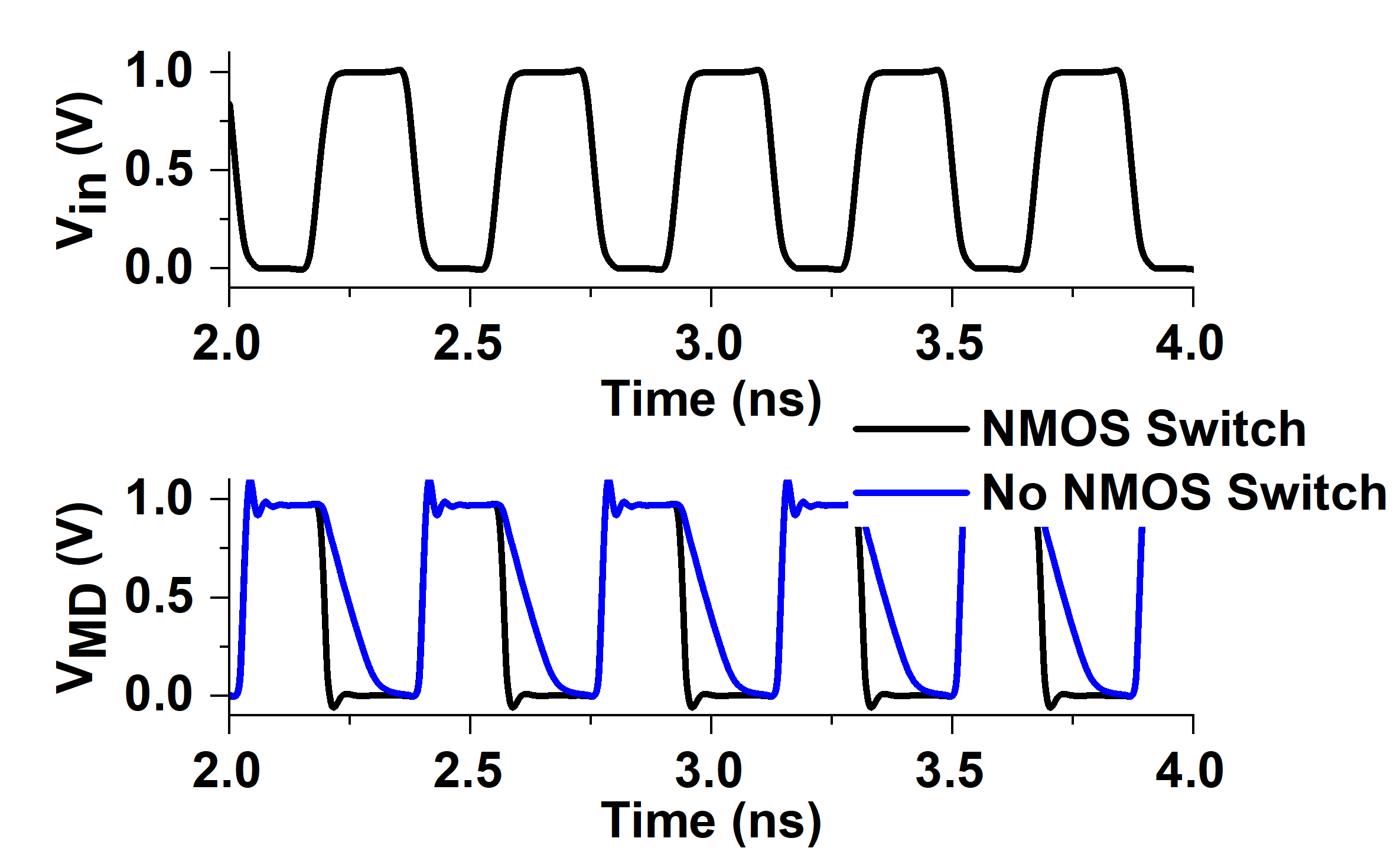}
\caption{Comparison the switching performance with and without extra NMOS switch. The first plot shows the input microwave signal ($V_{in}$) and the second one shows the voltage at the drain of the NMOS current source ($V_{MD}$).}\label{fig_switch_comp}
\end{figure}

The full schematic of the microwave launcher with the the switches and current sources banks is shown in Fig.~\ref{fig_array_circuits}. NMOS current mirrors are used to control the current in each individual wire of the array. PMOS switches are used to convert the output voltage of the PLL into microwave currents in the array. The current sources are design to enable current control from $I_0$ = 0.1~mA to $I_0$ = 1~mA. At low current values our simulation in Fig.~\ref{fig_switch_comp} indicates that when the PMOS switches are off, the current of wire returns to zero at a slow speed. This is due to the large parasitic capacitance of the wire array and the limited discharging current. To increase the pull down speed, a row of NMOS switches are also used (see Fig.~\ref{fig_array_circuits}) provide a fast discharging (see the comparison in Fig.~\ref{fig_switch_comp}). The transistors dimensions ($W_1$/$L_1$, $W_2$/$L_2$, and $W_3$/$L_3$) are scaled up by a factor of 3 at the boundaries. As described in Section~\ref{sec_launcher}, the microwave launcher is implemented on M8 with wire width and spacing of 2~$\upmu$m. Therefore, all the transistor layouts are designed in a multiplier fashion so that they can fit in the tight wire pitch. A pair of wide return ground paths are used to close the current loop.

Part of the photonic filter is also implemented on M8 with unit width and spacing of 400~nm (shown in Fig.~\ref{fig_three_layer_filter}). The smallest dimension limited by the design rule of the technology is selected to enhance the coupling to the surface plasmon polariton mode. To cater the needs of both components, shown in Fig.~\ref{fig_array_circuits}, each of the microwave launcher wires other than the boundary wires is divided into three sub-wires with their two ends electrically connected. The width and the spacing of these sub-wires are 400~nm. In addition, each of the 2~$\upmu$m spacing in the microwave launcher is filled with another two floating dummy wires with 400-nm width and spacing (Fig.~\ref{fig_array_circuits}). These dummy wires do not affect the microwave generation/delivery operation, but along with the other sub-wires, they form the desired phontonic filter geometry, which is 350$\times$130~$\upmu$m$^2$ of total area. In this prototype, a conservative size was adopted for the sensing area, since the edge effects like leakage of light was uncertain. The size of the photodiode placed under the filter is 300$\times$80~$\upmu$m$^2$. In future designs, the photodiode area may increase for better $SNR$.   

This array generates $>$95$\%$ field homogeneity over $>50\%$ of its area compared to 25$\%$ in \cite{Ibrahim2018}. Uniform array spacing is adopted in this design, however,  in Section~\ref{sec_conclusion}, we show that a certain gap between group~\emph{D} and group~\emph{A}/\emph{C} can achieve homogeneity across a larger area. The design is only enabled by CMOS technologies, where we can integrate a very tight wire array with efficient switching and current control.

\subsection{On-Chip Synthesis of the Microwave Frequency}\label{sec_PLL}

\begin{figure}[h]
\centering
\includegraphics[width=3.5in]{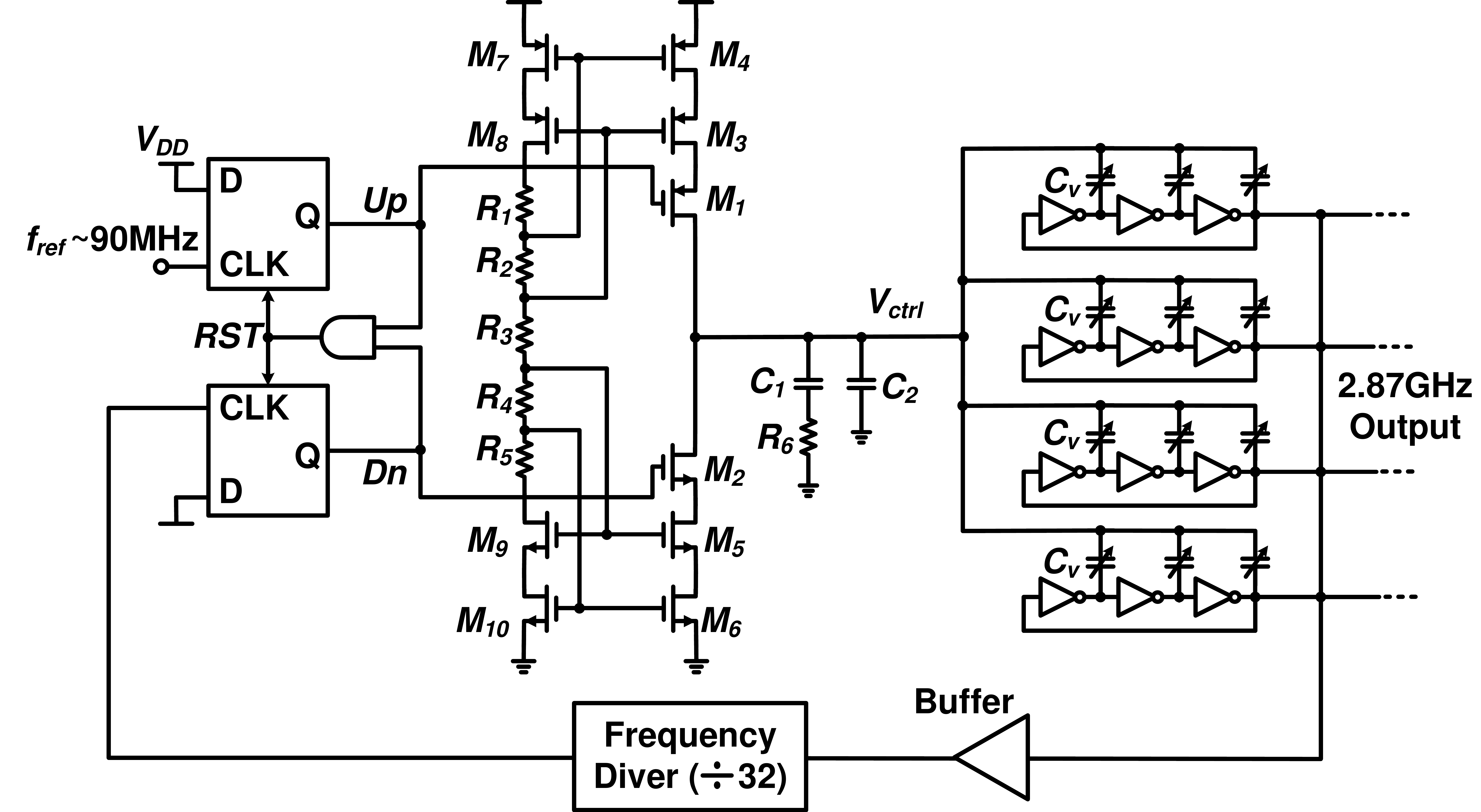}
\caption{Schematic of the 2.87-GHz PLL.}\label{fig_pll_schematic}
\end{figure}

A PLL loop is used to stabilize the output of the voltage-controlled oscillator (VCO) that drives the NV centers. The full schematic of the PLL is shown in Fig.~\ref{fig_pll_schematic}. To ensure a uniform phase of the magnetic field across the wire array, the VCO is based on a tightly-coupled ring-oscillator array. Four sub-oscillators are placed with a spacing of $\sim$50~$\upmu$m (see Fig.~\ref{fig_architecture}). The two oscillators in the middle drive 32 wires, while the other two oscillators on the array boundaries drive 8 wires each. Compared to a centralized signal-generation scheme, the coupled-oscillator eliminates the phase variation caused by different lengths of the microwave-distributing traces. Here, ring oscillators, instead of LC oscillators, are chosen, in order to eliminate the need for extra inductors, as well as any magnetic coupling and degradation of field homogeneity in the diamond-sensing area. A ring oscillator also offers wider tuning range and layout compactness. Its inferior phase noise is improved by 6~dB due to the oscillator coupling, and in fact our analysis indicates that our sensor sensitivity is currently limited by the green and red light noise, as explained in Section~\ref{sec_noise_analysis}. The VCO frequency is tuned using varactors ($C_v$=9~to~42~fF) with a coefficient of 1.2~GHz/V. The charge pump current of the PLL is 5~mA and the loop filter parameters are $C_1$=1~nF,~$R_6$=220~$\Omega$,~and~$C_2$=20~pF.

\section{Chip Prototype and Experimental Results}\label{sec_measurement}

\begin{figure}[h]
\centering
\includegraphics[width=3in]{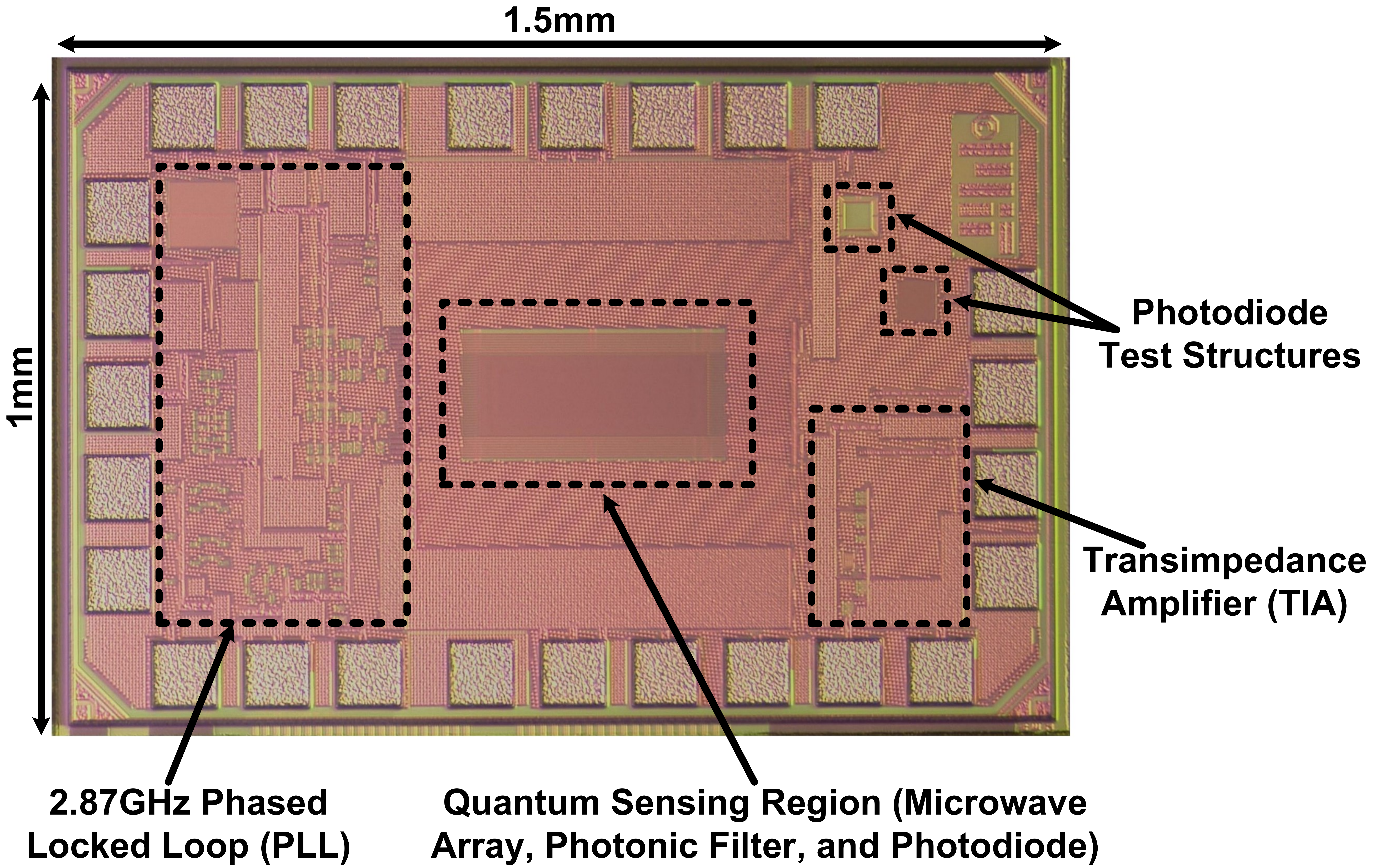}
\caption{The micrograph of the CMOS quantum magnetometry chip.}\label{fig_chip_photo}
\end{figure}

The chip prototype is fabricated using a TSMC 65-nm CMOS technology. The chip micrograph is shown in Fig.~\ref{fig_chip_photo}. The chip area has an area of 1$\times$1.5~mm$^2$ and consumes 40~mW of power. A diamond slab with NV centers is placed and attached onto the top of the CMOS chip as shown in Fig.~\ref{fig_architecture}.
This diamond is a single crystalline CVD-grown diamond from Element 6. It is electronically irradiated with a dosage of $10^{18} \ \text{e}^- /\text{cm}^{2}$ at 1~MeV, and then annealed for 2~hours at 850$^{\circ}$C. Immersion oil is used to adhere the diamond slab to the chip. By bridging the difference of the refractive index, the oil also minimizes the fluorescence loss. The measured filter rejection for green light (532~nm) is $\sim$25~dB, which well match the simulation results in Section~\ref{sec_filter}. A 45$^{\circ}$ cut is introduced in the diamond’s corner to direct the vertical incident green laser horizontally to further enhance the overall green rejection ratio to $\sim$33~dB. The measured responsivity of the on-chip photodiode is 0.19~A/W.

\begin{figure} [!b]
\centering
\subfloat[]{\includegraphics[width=3.3in]{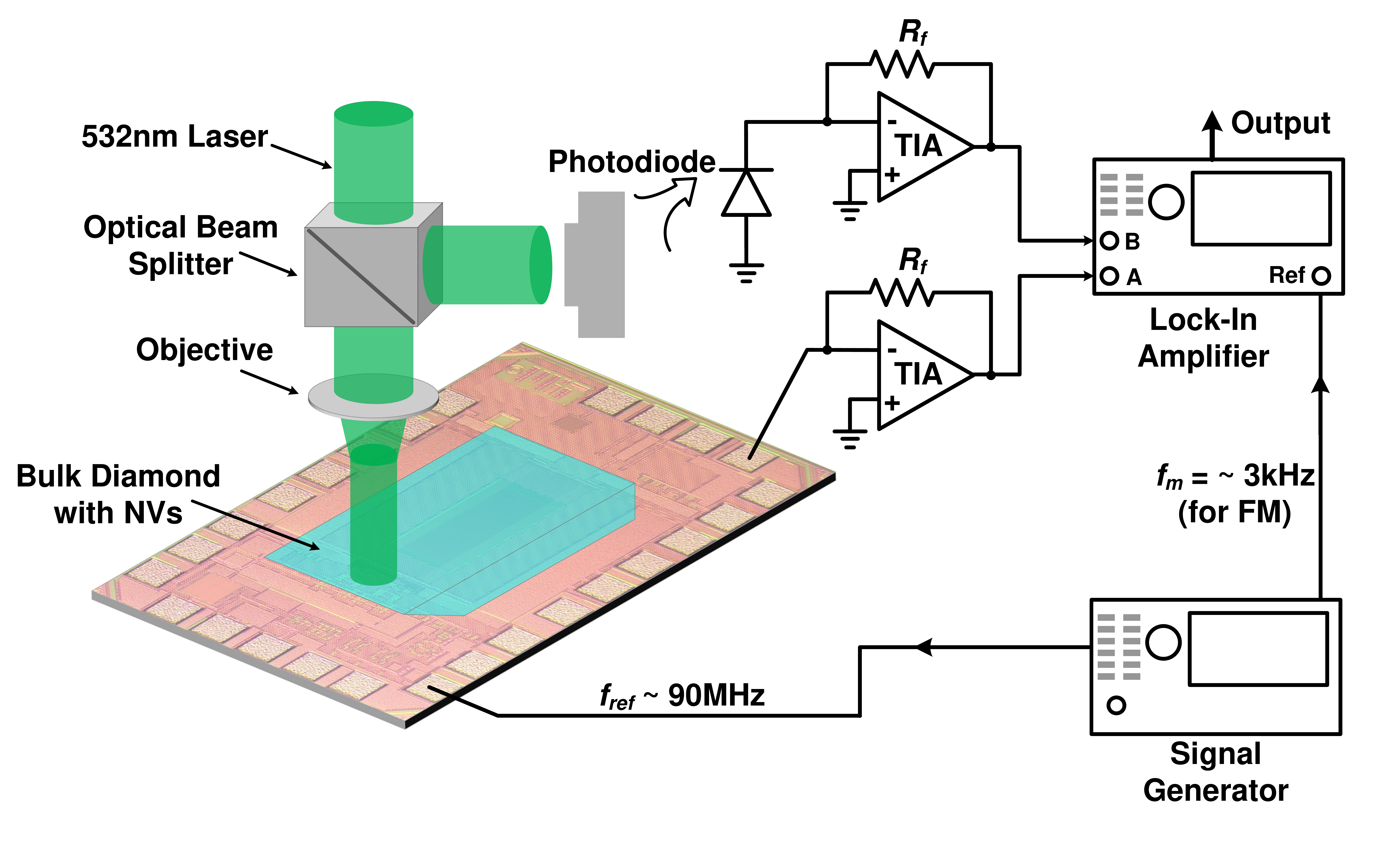}\label{fig_test_setup}}\\
\subfloat[]{\includegraphics[width=2.6in]{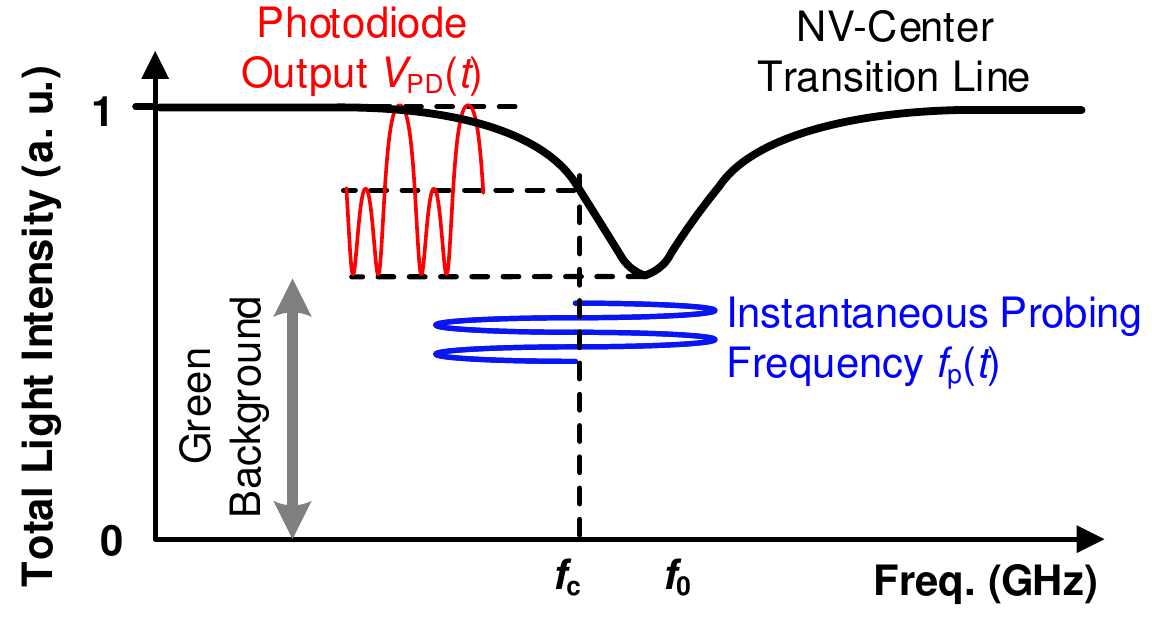}\label{fig_wms}}
  \caption{(a) The test setup for the hybrid NV-CMOS magnetometer. (b) The frequency-modulation scheme used in the setup.}
\end{figure}

\subsection{Optical Detected Magnetic Resonance Measurements}
The measurement setup to generate the ODMR spectral line is shown in Fig.~\ref{fig_test_setup}. A 532-nm laser light is used to excite the NV centers in the diamond. The red fluorescence from the on-chip photodiode is recorded while sweeping the reference signal ($\sim$100~MHz) of the on-chip PLL. Due to the limited green-red suppression ratio of the on-chip filter, the red fluorescence signal is still superposed on a large green-light background at the output of the photodiode. Meanwhile, we notice the green-light excitation exhibits large power fluctuation, so to remove the fluctuating background signal, a frequency-modulation (FM) scheme is used, where the chip-generated microwave probing frequency $f_p(t)$ is periodically varied with a deviation of 6~MHz and repetition frequency $f_{m}$ of 3~kHz. That is done by modulating the 100-MHz reference signal of the on-chip PLL. As shown in Fig.~\ref{fig_wms}, the corresponding photodiode output changes at $f_{m}$ (with harmonics at 2$f_m$, 3$f_m$, etc.). Such a time-varying component mainly results from the transition line and the associated red-fluorescence change, which is then readily measured by a lock-in amplifier (SR865) with the reference at $f_m$. The lock-in amplifier also provides narrow-band filtering around $f_m$. Note that the FM technique is common in spectroscopy\cite{Wang2018} and atomic/molecular clocks\cite{Wang2018b}. Lastly, to further de-embed the excitation noise within the lock-in amplifier bandwidth, a differential detection scheme is adopted, where a split beam of the same laser is measured by an off-chip photodiode; such a duplicated noise signal is then taken by the lock-in amplifier and is subtracted from the chip output. Two identical resistive transimpedance amplifiers (TIA) with gain of 10K are used to convert the photodiode currents into voltage. This gain is chosen to prevent the saturation of the amplifier due to the green current background (40~$\upmu$A)



\begin{figure}
\centering
\subfloat[]{\hspace{-4mm}\includegraphics[width=2.8in]{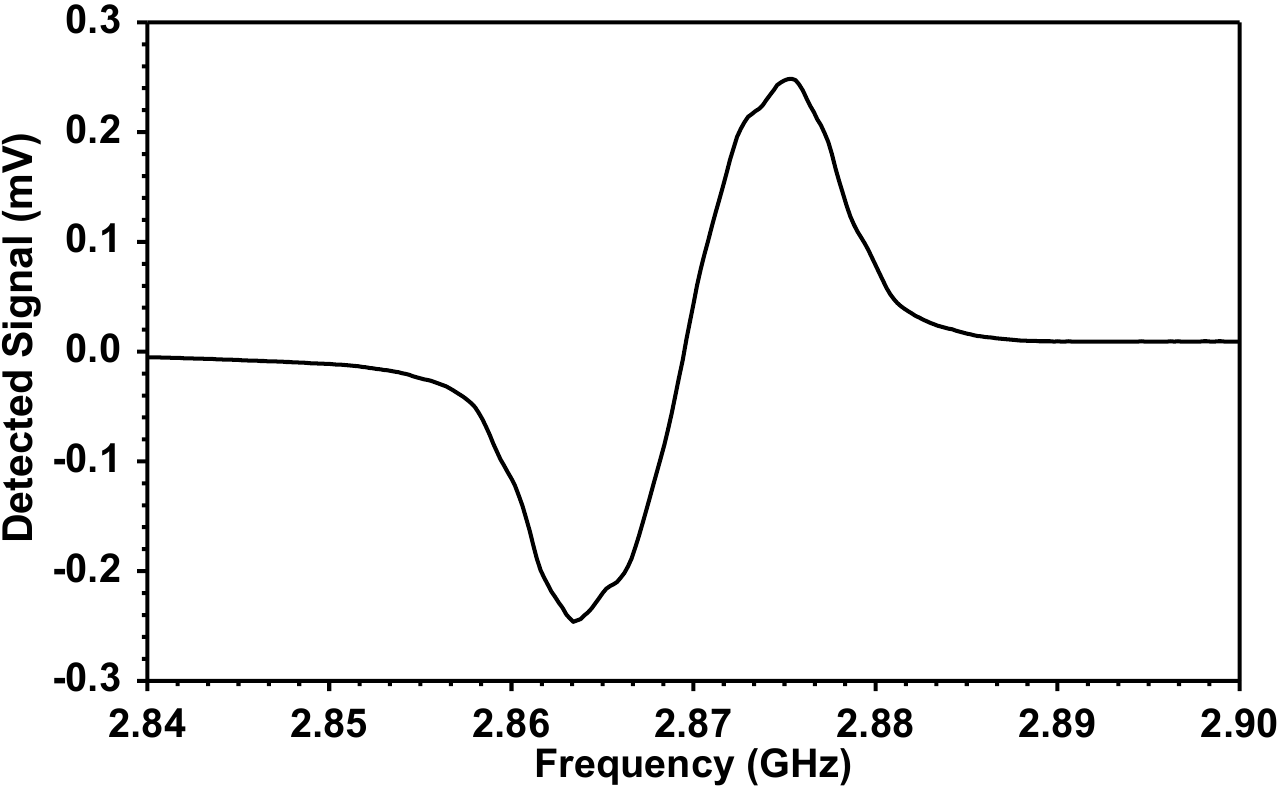}\label{fig_ODMR_NoField}}\\
\subfloat[]{\includegraphics[width=2.9in]{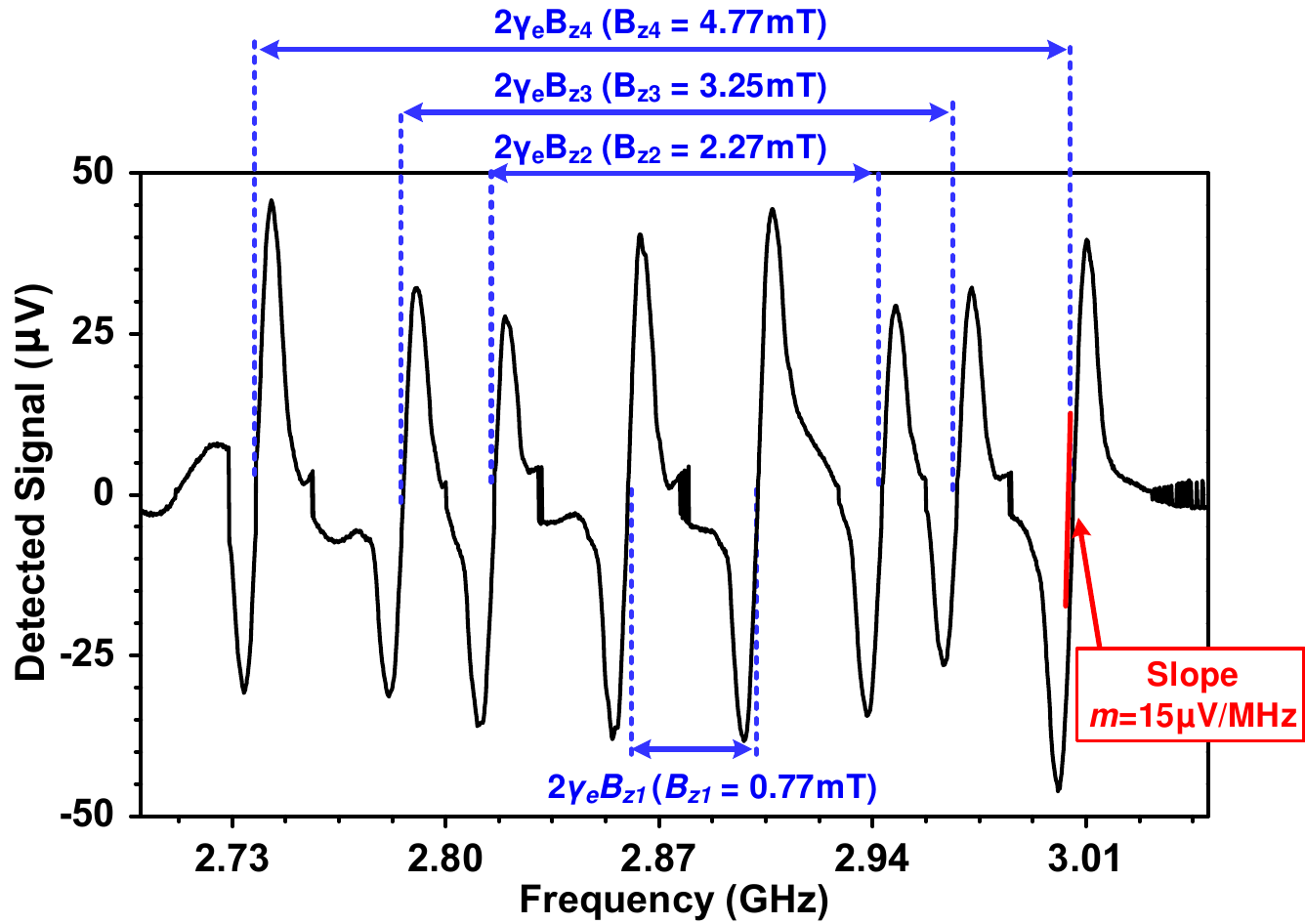}\label{fig_ODMR_Field}}
  \caption{Measured optically-detected magnetic resonance (ODMR) plot from the CMOS chip, when (a) no external magnetic filed is applied, and (b) a 5.4-mT external magnetic field is applied from a certain angle.}
\end{figure}

First, without an externally applied magnetic field, an ODMR plot is obtained as shown in Fig.~\ref{fig_ODMR_NoField}. A strong spin resonance, which results from all four NV centers in the diamond lattice, is detected at 2.87~GHz. The FM technique introduced previously can be interpreted as taking the first-order derivative of the regular Gaussian-shape transition line; as a result, the measured curve has the zigzag shape shown in Fig.~\ref{fig_ODMR_NoField}. The zero-crossing point corresponds to the actual transition frequency. Next, a 5.4-mT DC magnetic field is applied to the sensor from a nearby permanent magnet. This results in the Zeeman splitting for the spin states of the NV centers. As shown in the measured ODMR plot in Fig.~\ref{fig_ODMR_Field}, four pairs of resonances are observed. They correspond to the four NV-center orientations of the single crystalline diamond. The difference of the transition frequencies (i.e. zero-crossing points) in each resonance pair is proportional to the external static magnetic field along the associated NV-center orientation. The proportionality constant, as described in Section~\ref{sec_physics}, is $\gamma_e$=28~GHz/T. Accordingly, the magnetic fields along the four NV-center orientations are 0.77~mT, 2.27~mT, 3.25~mT, and 4.77~mT, respectively. Lastly, as the comparison between Fig.~\ref{fig_ODMR_NoField} and \ref{fig_ODMR_Field} shows, the chip output signal with external magnetic field is reduced due to the breaking of the degeneracy of the resonances from 2.87~GHz. Fig.~\ref{fig_ODMR_Field} clearly indicates that, by mapping the projections of the external magnetic field along the four NV-center orientations and certain frame of reference, one can construct the full vector magnetic field. 


One practical way to use the sensor for vector-field sensing, is to first ``bias" the sensor with a static magnetic field (e.g. a permanent magnet), which allows for the four resonances to be completely split (such as shown in Fig.~\ref{fig_ODMR_Field}). Then, we record the additional change of the sensor output voltage $\Delta v_{out,i}$ ($i$=$\pm$1, $\pm$2, $\pm$3, $\pm$4) around each zero-crossing point of the FM-based ODMR curve. Note that $\Delta v_{out,i}$ is caused by the shift of each resonance frequency $\Delta f_i$ due to the projection of the added field on the associated NV-center axis $\Delta B_{zi}$ (to be measured) on top of the bias static field, its expression is (for positive $i$):
\begin{align}\label{eqn_delta_vout}
    \Delta v_{out,i}=m\Delta f_i=m\gamma_e\Delta B_{zi}, 
\end{align}
where $m$ is the slope of the ODMR curve at each transition zero-crossing point ($\sim$15~$\upmu$V/MHz in our case, see Fig.~\ref{fig_ODMR_Field}). The value of $\Delta B_{zi}$ can be derived from $\Delta B_{zi}=\Delta v_{out,i}$/$(m\gamma_e)$. Note that the transition frequencies of both $\ket{m_s=\pm 1}$ sub-levels are also temperature dependent (with a coefficient of -74~kHz/K\cite{Acosta2010TemperatureDiamond}). To cancel such temperature-induced drifts, it is better to measure the differential change of each pair of $\ket{m_s=\pm 1}$ transitions, and use the following equation for $\Delta B_{zi}$\cite{Kim2019a}:
\begin{align}
    \Delta B_{zi}=\frac{\Delta v_{out,i}-\Delta v_{out,-i}}{2m\gamma_e}~~~(i=1, 2, 3, 4).
\end{align}


\subsection{Magnetic Sensitivity Estimation}\label{sec_noise_analysis}
\begin{figure}[t]
\centering
\includegraphics[width=3in]{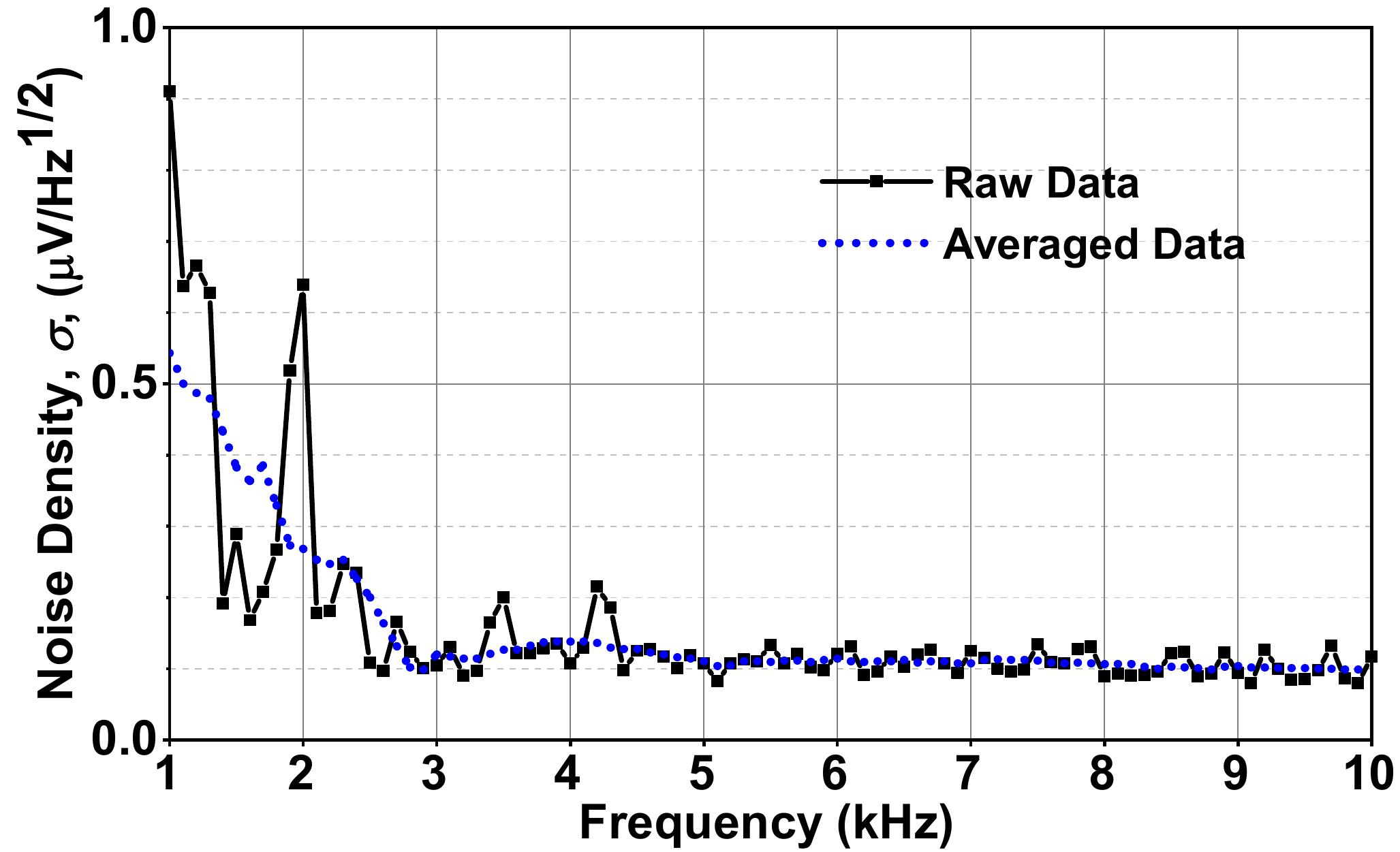}
\caption{The measured output noise floor of the sensor.}\label{fig_noise_floor}
\end{figure}

In order to calculate the magnetic sensitivity of the sensor, the noise floor, $\sigma$ (unit: V/Hz$^{1/2}$), is measured. The sensor noise is measured by monitoring the read-out of noise in the lock-in amplifier while sweeping the modulation frequency $f_m$. The results are shown in Fig.~\ref{fig_noise_floor}, where the sensor shot noise floor is $\sim$0.1~$\upmu$V/Hz$^{1/2}$. The increased noise below 3~kHz is caused by the interference from the unshielded testing environment. This allows us to calculate the magnetometer sensitivity $S$ (unit: T/Hz$^{1/2}$), which is the minimum detectable magnetic field with 1-s integration time, using the following equation:
\begin{equation}\label{eqn_sensitivity}
    S=\frac{\sigma}{\gamma_em}.
\end{equation}
The measured magnetometry sensitivity of this sensor at $f_m$=3~kHz is 245~nT/Hz$^{1/2}$.

The sensor performance is limited by the green light noise. The estimated theoretical magnetic sensitivity ($S_g$) of the sensor can be estimated using the following:

\begin{equation}\label{total_noise}
S_g = \frac{\sigma_g}{\gamma_e m} = \frac{\sqrt{2(2qi_gR_f^2 + 4KTR_f)}}{\gamma_e m}
\end{equation} 
where $\sigma_g$ is the theoretical noise limit from the green light. The first term of $\sigma_g$ represents the shot noise of the green light detected by the photodiode. $q$ is the electronic charge, $i_g$ is the current measured at the photodiode due to the green light ($\sim$40~$\upmu$A), and $R_f$ is the feedback resistor of the TIA (10~k$\Omega$). The second term represents the noise of the TIA (assuming it is mainly limited by $R_f$). $K$ is the Boltzmann's constant, $T$ is the temperature. The factor of 2 is due to the addition of the noise power due to the two detection branches. The calculated theoretical value is ~$\sim$150~nT/Hz$^{1/2}$, which agrees to good extent with the measured value, and the difference is due to the residual un-canceled green light fluctuation. Even though this differential measurements technique increase the total shot noise limit by a factor of $\sqrt{2}$, it helps in canceling almost all the external laser fluctuation, which is almost one order of magnitude larger than the shot noise limit in our case.

To confirm that the green light noise is the dominant source of noise. The contributions of the other noise sources in the system are estimated as follows:  \begin{enumerate}
     \item The magnetic sensitivity due to the NV red fluorescence shot noise ($S_r$) is $\sim 10~\text{nT} / \sqrt{\text{Hz}}$ at $f_m =3 \text{ kHz}$ as given in:
    
\begin{equation}\label{red_noise}
S_r = \frac{\sigma_r}{\gamma_e m}  = \frac{\sqrt{(2q i_r)R_f^2}}{\gamma_e m} \approx \frac{\sqrt{(2q\frac{V_{max}}{CR_f})R_f^2}}{\gamma_e m}.
\end{equation} 
In (\ref{red_noise}), $V_{max}$ is the maximum voltage of the ODMR curve in Fig.~\ref{fig_ODMR_Field}, which is $\sim$45~$\upmu$V, and $C$ is the ODMR contrast of $\sim 0.01$. 

     \item The magnetic sensitivity ($S_m$) due to the amplitude noise converted from the microwave generator spectral purity is determined by (see Appendix I):
     
     \begin{equation}
     \label{eqn_Sm}
     \begin{split}
     S_m = \frac{\sigma_m}{\gamma_e m} \approx \frac{2\pi f_m \phi_n m}{\gamma_e m} \approx \frac{2\pi f_m \phi_n}{\gamma_e},
     \end{split}
     \end{equation}
    where $m$ is the slope of the FM-ODMR curve. Based on a measured PLL phase noise ($\phi_n$) of -88~dBc/Hz at 3~kHz (FM modulation frequency $f_m$). The above magnetic sensitivity is $\sim 30~\text{pT}/\sqrt{\text{Hz}}$.


\end{enumerate}

\section{Conclusions}\label{sec_conclusion}

\begin{figure}[!b]
\centering
\includegraphics[width=3in]{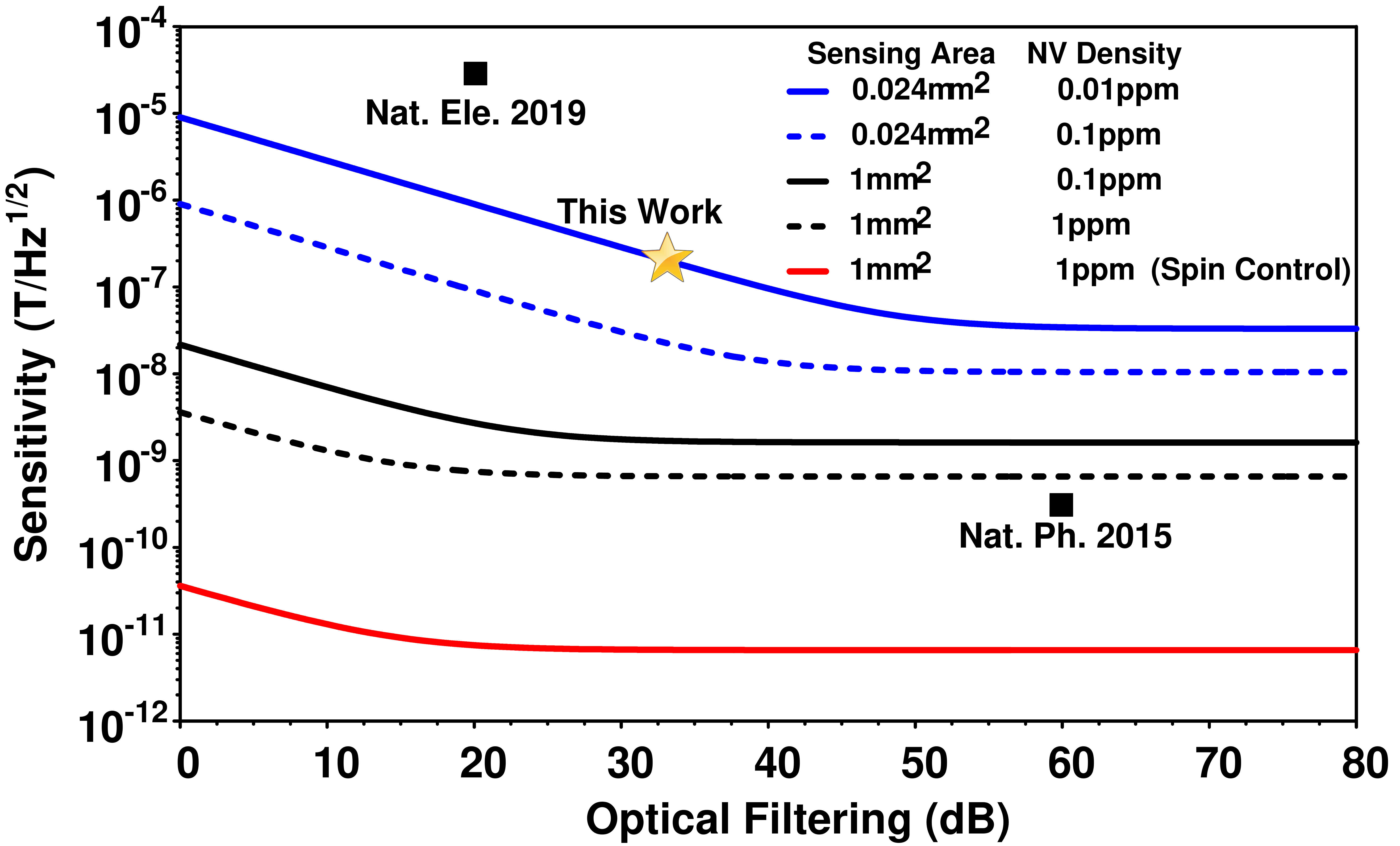}
\caption{Estimation of sensitivity for hybrid CMOS-NV magnetometers with different configurations.}\label{fig_sens_comp}
\end{figure}

In this paper, we present a vector-field magnetometer, which is based on a CMOS chip-scale platform with a hybrid integration of color centers (NV centers) in diamond. The chip consists of the essential components for manipulation and detection of the spin states of the NV centers in diamond. We demonstrate room-temperature magnetic field sensing with a sensitivity of 245~nT/Hz$^{1/2}$, which is 130$\times$ better than our previous prototype \cite{Kim2019a}. A microwave launcher with uniform microwave generation is introduced that can be scaled up for large diamond areas (i.e. large $SNR$). An enhanced multi-layer nanophotonic filter for lower background noise is also presented. 

The scalable architecture and component designs presented in this paper provide a clear pathway to further push the sensitivity of the proposed sensor to sub-nT/Hz$^{1/2}$. This is shown in Fig.~\ref{fig_sens_comp}, which plots the estimated sensitivity with varying optical filtering performance, sensing areas and NV-centers densities (see Appendix II for more details). As Shown in Fig.~\ref{fig_grating_patterns}, current locations for lower optical grating layers are still not optimal due to the fixed BEOL stack configuration. In more advanced CMOS technology nodes, where a denser stack of thinner metal layers are available, the optical filtering performance can be further improved. For the microwave launcher, a better optimized configuration of the boundary wire arrays (i.e. Group \emph{D} in Fig.~\ref{fig_finite_wire_array_w_boundary}) can lead to high magnetic homogeneity over a larger proportion of the total launcher area. For example, we find that if each Group~\emph{D} consists of $m$=4 wires ($d$=4~$\upmu$m) with $I_D$=6.35~mA, and a gap of 12~$\upmu$m exists between Group~\emph{D} and Group~\emph{A}/\emph{C}, 95\% magnetic homogeneity is achieved for the entire -60 to 60-$\upmu$m space above Group \emph{A-B-C}. Homogeneity over large space also promises a significant sensitivity enhancement via spin-controlled pulse sequence \cite{barry2019sensitivity}. Lastly, the ability to manipulate the NV centers over a larger area also enables the integration of larger numbers of detectors for gradient magnetometry, multiplexed analytical nuclear-magnetic resonance (NMR) spectroscopy\cite{glenn2018high}, atomic gyroscopes\cite{Jaskula_2019}, and other quantum-sensing applications \cite{degen2017quantum}.


\section*{Acknowledgment}
The authors would like to thank Xiang Yi, Donggyu Kim, David Bono, and Cheng Peng at MIT for technical discussions and their assistance during prototyping and testing.
\section*{Appendix I: Estimation of the Contribution of the Microwave Phase Noise on the Sensor Sensitivity}\label{sec_appendix_2}

The microwave signal that drives the NV centers can be represented as:\begin{equation}
\label{eqn_VRF}
\begin{split}
    V_{RF} = &V_0 \cos{(\omega_0t+\phi_n(t))},
\end{split}
\end{equation}
where $V_0$, and $\omega_0$ are the microwave signal amplitude and frequency, respectively. 
$\phi_n(t)$ is the phase noise of the microwave frequency. The instantaneous frequency can be represented as: 
\begin{equation}
\label{eqn_Omegat}
\begin{split}
    \omega(t)= \omega_0+\frac{d\phi_n(t)}{dt},
\end{split}
\end{equation}
The frequency fluctuations around the operating frequency ($\omega_0$) is related to the phase noise by:
\begin{equation}
\label{eqn_deltaf}
\begin{split}
    \delta\omega = \frac{d\phi_n(t)}{dt},
\end{split}
\end{equation}
In our case, a lock-in amplifier is used to observe the noise around the modulation frequency ($\omega_m$). Therefore, $\phi_n(t)$=$\phi_n\cos{2\pi f_m t}$ can be assumed to have a sinusoidal form, where $\phi_n$ is the phase noise of the microwave source at the modulation frequency ($f_m$) in dBc/Hz. This means the peak frequency fluctuation is $\approx 2\pi f_m \phi_n$. Therefore the voltage spectral density due to the PM to AM noise conversion is represented as: \begin{equation}
\label{eqn_Vn}
\begin{split}
    \sqrt{\frac{\bar{V_n^2}}{\delta f}} \approx  2\pi f_m \phi_n m,
\end{split}
\end{equation}
where $m$ is the slope of the FM curve in V/Hz.




\section*{Appendix II: Estimation of Sensitivity for Future Hybrid CMOS-NV Magnetometers}\label{sec_appendix}


The sensor sensitivity is inversely proportional to the $SNR$ of the sensor output:

\begin{equation}\label{eqn_sensitivity_App}
    S \propto \frac{1}{SNR} = \frac{\sigma_i\cdot\Delta\nu}{\gamma_e\cdot I_r},
\end{equation}
where $\sigma_i$ is the total noise current density (unit: A/Hz$^{1/2}$), $\Delta\nu$ is the linewidth of the ODMR curve, $I_r$ is the photo-current due to the red fluorescence signal, and $\gamma_e$ is 28~GHz/T. From our experiments, the linewidth of the ODMR is 6~MHz. The noise is limited by the shot noise of either the green excitation (for low optical filtering ratio) or the red fluorescence (for high optical filtering ratio): 

\begin{equation}\label{eqn_noise_App}
    \sigma_i = \sqrt{2\cdot q \cdot (I_g + I_r)},
\end{equation}
where $q$ is the electronic charge, and $I_g$ and $I_r$ are the photo-currents due to the unfiltered green light and red fluorescence, respectively. The value of $I_g$ is estimated as:
\begin{equation}\label{eqn_Igreen_App}
    I_g = P_g\cdot R_{PD}\cdot\eta_g,
\end{equation}
where $P_g$ is the input optical power, $R_{PD}$ is the photodiode responsivity, and $\eta_g$ is the green rejection of the optical filter. The value of $I_r$ is proportional to the total number $n_{nv}$ of NV centers in the diamond, which is proportional to the sensing area $A_{nv}$ and NV density $\rho_{nv}$. Lastly, the performance using the spin-controlled pulse sequence is estimated based on the fact that the equivalent transition linewidth (hence the zero-crossing slope $m$) of the configuration is reduced by $\sim$100$\times$ for AC magnetic field measurements \cite{Taylor2008High-SensitivityResolution}, \cite{barry2019sensitivity}.



\bibliographystyle{IEEEtran}
\bibliography{main.bib}


\end{document}